\documentclass[11pt]{arxiv}
\pdfoutput=1
\usepackage{etoolbox}
\usepackage[linesnumbered,ruled]{algorithm2e}
\usepackage{algpseudocode}
\usepackage{bbm}
\usepackage{epstopdf}
\usepackage{tikz}
\usepackage{authblk}

\usepackage{enumitem}


\usetikzlibrary{shapes,arrows,positioning}
\usetikzlibrary{calc}
\usetikzlibrary{patterns,decorations.pathreplacing}

\let\oldnl\nl
\newcommand{\nonl}{\renewcommand{\nl}{\let\nl\oldnl}}

\ifcsdef{spenumerate}{}{
\newenvironment{spenumerate}{%
   \begin{list}{(\arabic{enumi})}{%
    \setlength\labelwidth{3.5em}%
    \setlength\leftmargin{2.5em}%
    \setlength{\topsep}{2pt plus 2pt minus 2pt}%
    \setlength\itemsep{0.0cm}%
    \usecounter{enumi}}%
  }{\end{list}}
}

\graphicspath{{./Figures/}}
\begin{document}

\title{Better Algorithms for Hybrid Circuit and Packet Switching in Data Centers}

\author{Liang Liu\thanks{lliu315@gatech.edu}}
\author{Long Gong\thanks{gonglong@gatech.edu}}
\author{Sen Yang\thanks{sen.yang@gatech.edu}}
\author{Jun (Jim) Xu\thanks{jx@cc.gatech.edu}}
\author{Lance Fortnow\thanks{fortnow@gatech.edu}}
\affil{Georgia Institute of Technology}
%
%
%
%
%
%
%
\maketitle
\begin{abstract}

Hybrid circuit and packet switching for data center networking (DCN)
has received considerable research attention recently.
A hybrid-switched DCN employs a much faster
circuit switch that is reconfigurable with a nontrivial cost, and a much slower packet switch that is reconfigurable with no cost,
to interconnect its racks of servers.  The research problem is, given a traffic demand matrix (between the racks),
how to compute a good circuit switch configuration schedule so that the vast majority of the traffic demand is removed by the circuit switch, leaving a
remaining demand matrix that contains only small elements
for the packet switch to handle.   In this paper, we propose two new hybrid switch scheduling algorithms under two different scheduling constraints.
Our first algorithm, called 2-hop Eclipse,
strikes a much better tradeoff between the resulting performance (of the hybrid switch) and the computational complexity (of the algorithm)
than the state of the art solution Eclipse/Eclipse++.   Our second algorithm, called BFF (best first fit), is the first hybrid switching solution that exploits
the potential partial reconfiguration capability of the circuit switch for performance gains.

\end{abstract}

\section{Introduction}

Fueled by the phenomenal growth of cloud computing services,
data center network continues to grow relentlessly both in size, as measured by the number of racks of servers it has to interconnect, and in speed, as measured by the amount of traffic it has to
transport per unit of time from/to each rack~\cite{decusatis2014optical}.
A traditional data center network (DCN)
architecture typically consists of a three-level multi-rooted tree of switches that start, at the lowest level, with the Top-of-Rack (ToR) switches,
that each connects a rack of servers to the network \cite{patel2013ananta}.
However, such an architecture has become increasingly unable to scale with the explosive growth in both the size and the speed of the DCN, as we can no longer increase the transporting and switching
capabilities of the underlying commodity packet switches without increasing their costs significantly.


\subsection{Problem Statement}\label{sec:problem_statement}

\begin{figure}
\centering
\includegraphics[width=2.8in]{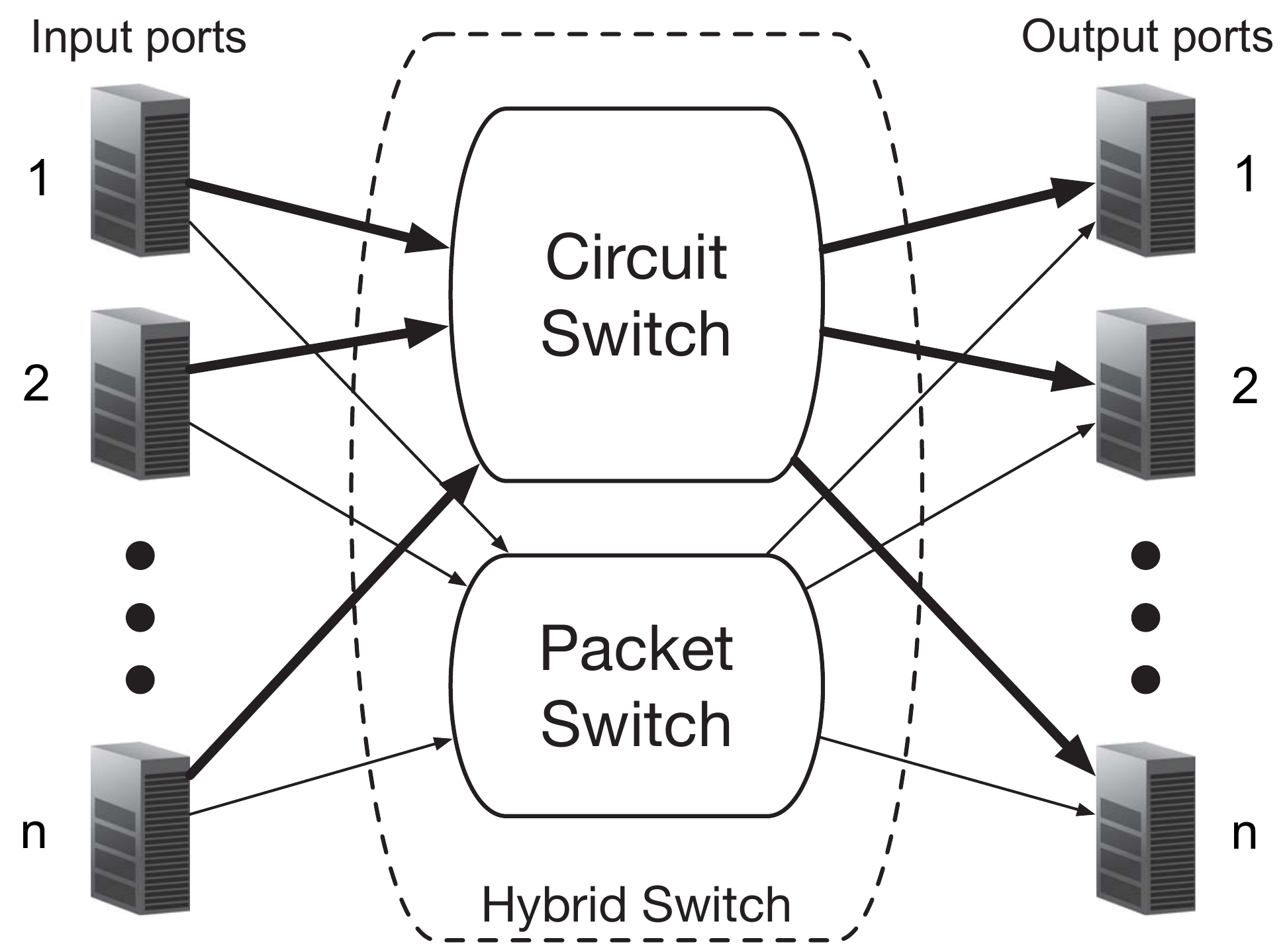}
\caption{Hybrid Circuit and Packet Switch}
\label{fig:hybrid_switch}
\end{figure}

A cost-effective solution approach to this scalability problem, called hybrid DCN architecture, has received considerable research attention in recent years~\cite{farrington2010helios,wang2010design,farrington201310}.
In a hybrid data center, shown in Figure~\ref{fig:hybrid_switch}, $n$ racks of computers on the left hand side (LHS)
are connected by both a circuit switch and a packet switch to $n$ racks on the right hand side (RHS).   Note that racks
on the LHS is an identical copy of those on the RHS;  however we restrict the role of the former to only transmitting data and
refer to them as {\it input ports}, and restrict the role of the latter to only receiving data and refer to them as {\it output ports}.
This purpose of this duplication (of racks) and role restrictions is that the resulting hybrid data center topology can be modeled as a
bipartite graph.

Each switch transmits data from input ports (racks on the LHS) to output ports (racks on the RHS)
according to the configuration of the switch at the moment.
Each switch is modeled as a bipartite graph, and its configuration
a bipartite matching over this bipartite graph.
The circuit switch has a much higher bandwidth than the packet switch (typically an order of magnitude higher), but incurs a nontrivial
reconfiguration delay $\delta$ when the switch configuration (\textit{i.e.}, the matching) has to change.   Depending on the underlying technology
of the circuit switch, $\delta$ can range from tens of microseconds to tens of milliseconds \cite{farrington2010helios,chen2014osa,wang2010c,liu2014circuit,Porter2013TMS}.

In this paper, we study an important optimization problem stemming from hybrid circuit and packet switching:  Given a traffic demand matrix
$D$ from input ports to output ports,
how to schedule the circuit switch to best (\textit{i.e.}, in the shortest amount of total transmission time or equivalently with the highest throughput)
meet the demand?
A schedule for the circuit switch consists of a sequence of configurations (matchings) and their time durations
$(M_1, \alpha_1)$, $(M_2, \alpha_2)$, $\cdots$,
$(M_K, \alpha_K)$.
A workable schedule should let the circuit switch remove (\textit{i.e.}, transmit)
the vast majority of the traffic demand from $D$, so that every row or column sum of the remaining demand matrix is small enough for the packet switch to handle.
Since the problem of computing the optimal schedule for hybrid switching, in various forms, is NP-hard~\cite{li2003scheduling}, almost all existing solutions
are greedy heuristics.
In this paper, we propose two greedy heuristic solutions to this problem under different scheduling constraints.

\subsection{State of the Art:  Eclipse and Eclipse++}\label{sec:state_of_art}

Since one of these two solutions builds upon the state of the art solution called Eclipse~\cite{bojjacostly},
we provide here a brief description of it, and its companion algorithm Eclipse++~\cite{bojjacostly}.
Eclipse iteratively chooses the aforementioned sequence of circuit switch configurations, one per iteration,
according to the following greedy criteria:  In each iteration, Eclipse tries to extract and subtract a matching (with its duration) from the $n\times n$ traffic demand matrix $D$ that has the largest {\it cost-adjusted utility}, which we will specify precisely in \autoref{sec:overview_eclipse}. Eclipse, like most other hybrid switching algorithms, considers and allows only direct routing in the following sense:  All circuit-switched data packets reach
their respective final destinations in one-hop (\textit{i.e.}, enters and exits the circuit switch only once).



However, restricting the solution strategy space to only direct routing algorithms may leave the circuit switch underutilized.
For example, a connection (edge) from input $i_0$ to output $j_0$ belongs to a matching $M$ that lasts $50$ microseconds, but at the start time of this matching,
there is only $40$ microseconds worth of traffic left for transmission from $i_0$ to $j_0$,
leaving $10$ microseconds of ``slack'' (\textit{i.e.}, \textit{residue capacity}) along this connection.
The existence of connections (edges) with
such ``slacks'' makes it possible to perform indirect (\textit{i.e.,} multi-hop) routing of remaining traffic via one or more relay nodes through a path
consisting of such edges.



Besides Albedo~\cite{li2017using}, Eclipse++~\cite{bojjacostly} is the only other work that has explored indirect routing in hybrid switching.
It was shown in~\cite{bojjacostly} that optimal indirect routing using such ``slacks'', left over by a direct routing solution such as Eclipse, can be
formulated as maximum multi-commodity flow over a ``slack graph'', which is itself
NP-complete~\cite{even1975complexity,ford1958suggested,hu1963multi,garg2007faster}.
Eclipse++ is a greedy heuristic solution that converts, with ``precision loss'' (otherwise P = NP), this multi-commodity flow computation to a large set of shortest-path computations.
Hence the computational complexity of Eclipse++ is still extremely high:
Both us and the authors of~\cite{bojjacostly} found that Eclipse++ is roughly three orders of magnitude more computationally expensive than Eclipse~\cite{personal} for a data center with $n = 100$ racks.
It was shown in~\cite{bojjacostly} that Eclipse++ can switch roughly $10\%$ more data, via indirect routing, than Eclipse under representative workloads.

\subsection{Our Solutions}
\label{sec:intro-solutions}

Our first algorithm, called 2-hop Eclipse, builds upon and significantly improves the performance of Eclipse~\cite{bojjacostly}.  Its design is based on the following subtle insight: Eclipse, the greedy heuristic,
\textit{w.r.t.} the aforementioned cost-adjusted utility function, to the original optimization problem that allows only direct routing, can be converted into a greedy heuristic, \textit{w.r.t.} the same utility function,
to a new optimization problem that allows both direct and 2-hop indirect routing.  In other words, 2-hop Eclipse is a slight, but very subtle, modification of Eclipse.  Hence, its asymptotical computational complexity is the same as, and its execution time only slightly higher than, that of Eclipse.  Yet, given the same workloads (the traffic demand matrices) used in~\cite{bojjacostly}, 2-hop Eclipse harvests, from indirect routing, much more performance gain (over Eclipse) than Eclipse++, which is three orders of magnitude more computationally expensive.


We emphasize that 2-hop Eclipse is very different than Eclipse++.  In particular, it has nothing to do with ``Eclipse++ restricted to 2-hops'', which according to our simulations performs slightly worse than,
and has almost the same computational complexity as, (unrestricted) Eclipse++. 
There are three main differences between 2-hop Eclipse and Eclipse++.  First, the relationship between Eclipse++ and Eclipse is ``arm's length'':
The only linkage between them is that the aforementioned ``slack graph'', which is a part of the output from Eclipse, is the input to Eclipse++.  That between 2-hop Eclipse and Eclipse, in contrast, is ``intimate'':  The former is
a slight modification of the latter.  Second, 2-hop Eclipse and Eclipse++ are greedy heuristic solutions to two very different NP-complete problems, as explained earlier.  Third, 2-hop Eclipse,
in each step (iteration) of its algorithm, performs a joint optimization of both direct and indirect routing, whereas Eclipse++ optimizes only indirect routing, over the aforementioned ``slack graph'' left over by Eclipse, after Eclipse is done optimizing direct routing.  This difference best explains the significant outperformance of 2-hop Eclipse over Eclipse++.

Our second algorithm solves this hybrid switching problem under a different scheduling constraint.
All existing works on hybrid switching make the convenient assumption
that when the circuit switch changes from the current matching $M_k$ to the next matching
$M_{k+1}$, every edge in $M_k$ has to stop data transmission during the reconfiguration period (of duration $\delta$),
even if the same connection continues on into $M_{k+1}$.  This is however an outdated and unnecessarily restrictive
assumption because all electronics or optical
technologies underlying the circuit switch can readily support {\it partial reconfiguration}
in the following sense:  Only the input ports affected by the configuration change need to pay a reconfiguration delay $\delta$,
while unaffected input ports can continue to transmit data during the configuration change.  For example, in cases where free-space
optics is used as the underlying technology (\textit{e.g.}, in~\cite{Ghobadi2016projector,Hamedazimi2014FireFly}), only each input port
affected by the configuration change needs (to rotate its micro-mirror) to redirect its laser beam towards its new output port and
incur the reconfiguration delay.

Our second algorithm, called Best First Fit (BFF), is the first hybrid switching solution that exploits the partial reconfiguration capability
for performance gains.  Unlike 2-hop eclipse, BFF considers only direct routing.
However, as we will show in \autoref{sec:evaluation}, the performance of BFF is either similar to or better than
that of 2-hop Eclipse, but the
computational complexity of BFF is roughly three orders of magnitude smaller than that of 2-hop Eclipse.  In other words, with this partial reconfiguration capability,
the hybrid switch can ``work much less hard'' to arrive at a schedule that is just as good or even better.






The rest of the paper is organized as follows. In \autoref{sec:model}, we describe the system model and the design objective this hybrid switch scheduling problem
in details.  In \autoref{sec:overview_eclipse}, we provides a more detailed description of Eclipse~\cite{bojjacostly}.
In \autoref{sec:indirect_routing} and \autoref{sec:partial_reconfig2}, we present our two solutions, 2-hop Eclipse and BFF, respectively.
In \autoref{sec:evaluation}, we evaluate the performance of our solutions against Eclipse and Eclipse++.
Finally, we describe related work in~\autoref{sec: related-work} and conclude the paper in~\autoref{sec:conclusion}.

\section{System Model and Problem Statement}\label{sec:model}

In this section, we present the system model and the problem statement of hybrid circuit and packet switching.


\subsection{System Model}

In \autoref{sec:problem_statement} and \autoref{fig:hybrid_switch}, we have already briefly introduced the system model for hybrid circuit and packet switching.
Here we expand a bit more on the subject.
As explained earlier, the circuit switch is modeled as a crossbar, so at any moment of time,
each input port can communicate with at most one output port, and vice versa, as dictated by the crossbar configuration,
a bipartite matching, at the time.   Borrowing the term virtual output queue (VOQ) from
the crossbar switching literature, we refer to, the set of packets that arrive at input port $i$ and are destined for output $j$,
as VOQ$(i, j)$.  In some places, we refer to a VOQ also as an input-output flow.

In a hybrid switching system, the circuit switch is usually an optical switch~\cite{farrington2010helios,chen2014osa,wang2010c},
and the packet switch is an electronic switch.  Hence the circuit switch is typically an order of magnitude or more faster
than the packet switch, as mentioned earlier.   For example, the circuit and packet switches might operate at the respective rates of
100 Gbps and 10 Gbps per port.  As mentioned earlier, the circuit switch
imposes a reconfiguration delay $\delta$ whenever its crossbar configuration changes,
whereas the packet switch does not.

For the time being,
we make the restrictive assumption that the circuit switch is not partially reconfigurable.   In other words,
during the reconfiguration period, no
input port can transmit any data.  We will lift this restriction in~\autoref{sec:partial_reconfig2}, where we present our BFF algorithm
that makes use of the partial reconfiguration capability for further performance gains.

\subsection{Traffic Demand and Batch Scheduling}

The demand matrix entry $D(i, j)$ is the amount of VOQ$(i, j)$ traffic, within a scheduling window, that needs to be
scheduled for transmission by the hybrid switch.
It was {\it effectively} assumed, in all prior works on hybrid switching except Albedo~\cite{li2017using} (to be discussed in \autoref{subsec: hybrid-switch}),
that the demand matrix $D$ is precisely known
before the computation of the circuit switch schedule begins (say at time $t$).
This assumption certainly makes sense if the hybrid switch is batch-scheduling
only ``inventory'' traffic, \textit{i.e.}, those packets that have arrived (before $t$).
However, to further reduce the delay to the network traffic, the scheduling window should ideally be allowed to end after $t$.
In other words, the demand matrix $D$ could also include future (\textit{w.r.t.} the time $t$) traffic.
It was stated in most of these prior works that their algorithms can accommodate such future traffic by (accurately)
forecasting their amounts, but how this forecasting is done is ``orthogonal'' to these algorithms.
In this work, we adopt the same {\it effective} assumption that $D$ is precisely known.

All prior hybrid switching algorithms except Albedo~\cite{li2017using} perform only batch scheduling (of ``inventory'' traffic or
``accurately forecasted'' future traffic).
In other words, given a demand matrix $D$, the schedules of the circuit and the packet switches are computed
before the transmissions of the batch (\textit{i.e.}, traffic in $D$) actually happen.  Our 2-hop Eclipse and BFF algorithms
are also designed for batch scheduling only.
Since batch scheduling is offline in nature (\textit{i.e.}, requires no irrevocable online decision-making), both 2-hop Eclipse and BFF algorithms
are allowed to ``travel back in time'' and modify the schedules of the packet and the circuit switches, and both will need to.



\subsection{Problem Formulation}\label{sec:scheduling_problem}

In this work, we study this problem of hybrid switch scheduling under the following standard formulation that was introduced
in~\cite{liu2015scheduling}:
to minimize the amount of time for the circuit and the packet switches working together to transmit a given traffic demand matrix $D$.
We refer to this amount
of time as {\it transmission time} throughout this paper.  An
alternative formulation, used in~\cite{bojjacostly}, is to maximize the amount of traffic that the hybrid switch can transmit within
a scheduling window of a fixed duration.  These two formulations are roughly equivalent, as mathematically
the latter is roughly the dual of the former.


A schedule of the circuit switch consists of a sequence of circuit switch configurations and their durations: $(M_1, \alpha_1)$, $(M_2, \alpha_2)$, $\cdots$, $(M_K, \alpha_K)$.
Each $M_k$ is an $n\times n$ permutation (matching) matrix;
$M_k(i,j) = 1$ if input $i$ is connected to output $j$ and $M_k(i,j) = 0$ otherwise.
We use the above notations for circuit switch schedules throughout this paper except in~\autoref{sec:partial_reconfig2}, where the solution based on the partial reconfiguration capability calls for a very different representation.
The total transmission time of the above schedule is $K \delta+\sum_{k=1}^K \alpha_k$, where $\delta$ is the reconfiguration delay.

Since computing the optimal circuit switch schedule {\it alone} (\textit{i.e.}, when there is no packet switch), in its full generality, is NP-hard~\cite{li2003scheduling},
almost all existing solutions
are greedy heuristics.  Indeed, the typical workloads
we see in data centers exhibit two characteristics that are favorable to such greedy heuristics:  sparsity (the vast majority of the demand matrix
elements have value $0$ or close to $0$) and skewness (few large elements in a row or column account for the majority of the row or column
sum)~\cite{liu2015scheduling}.

\section{Overview of Eclipse}\label{sec:overview_eclipse}




Since our 2-hop Eclipse algorithm builds upon Eclipse~\cite{bojjacostly}, we provide here a more detailed description of Eclipse (in \autoref{sec:overview_eclipse}).
Eclipse iteratively chooses a sequence of configurations, one per iteration,
according to the following greedy criteria:  In each iteration, Eclipse tries to extract and subtract a matching from the demand matrix $D$ that has the largest
{\it cost-adjusted utility}, defined as follows.
For a configuration $(M,\alpha)$ (using a matching $M$ for a duration of $\alpha$), its utility $U(M,\alpha)$, before adjusting for cost, is
$U(M,\alpha) \triangleq \|\min(\alpha M,D_{\mbox{\footnotesize rem}})\|_1$. Here $M$ is an $n\times n$ permutation matrix that denotes the configuration and $\alpha$ denotes its duration (defined in \autoref{sec:scheduling_problem}). $D_{\mbox{\footnotesize rem}}$ denotes what remains of the traffic demand (matrix) after we subtract from $D$
the amounts of traffic to be served by the circuit switch according to the previous matchings, \textit{i.e.}, those computed in the the previous iterations.
The $\|\cdot\|_1$ denotes the entrywise $L_1$ matrix norm, which is the summation of the absolute values of all matrix elements. In other words,
$U(M,\alpha) \triangleq \sum_{(i, j): M(i, j) = 1} \min\big{(}\alpha, D_{\mbox{\footnotesize rem}}(i, j)\big{)}$.
The cost of the configuration is modeled as $\delta + \alpha$, which accounts for the reconfiguration delay $\delta$.
The cost-adjusted utility is simply their quotient $\frac{U(M,\alpha)}{\delta + \alpha}$.  



Note that, given a fixed $\alpha$, optimizing the nominator $U(M,\alpha)$
alone is the \textit{maximum weighted matching problem} (MWM).
With the denominator $\delta + \alpha$ added
to the picture and $\alpha$ becoming an optimization variable, this optimization problem looks daunting.
Fortunately, it was proved in~\cite{bojjacostly} that the maximum value of the cost-adjusted utility is attained where $\alpha$ is equal to one of the matrix elements (\textit{i.e.}, lies on the support of $D_{\mbox{\footnotesize rem}}$).
However, performing a MWM computation for each of the $O(n^2)$ nonzero elements of the matrix $D_{\mbox{\footnotesize rem}}$ is computationally expensive.
Again fortunately, it was shown in~\cite{bojjacostly} that a binary search over these nonzero elements for a suitable $\alpha$ value results can result in a $(M,\alpha)$ that allow its cost-adjusted utility to attain the local maximum and
that together with such matchings thus in other iterations, results in good throughput performance empirically.  With this binary search, $O(4\log n)$ instead of $O(n^2)$ MWM computations are performed in each iteration.
Each MWM computation has complexity $n^{5/2} \log B$, where $B$ is the value of the largest element in $D$, using the scaling algorithm proposed in \cite{duan2012scaling}.
Hence the computational complexity of Eclipse is $O(Kn^{5/2}\log n\log B)$, shown in Table \ref{tab:complexity}, where $K$ is the total number of matchings (iterations).

\section{2-hop Eclipse}\label{sec:indirect_routing}

Unlike Eclipse, which considers only direct routing, 2-hop Eclipse considers both direct routing and 2-hop indirect routing in its optimization.
More specifically, 2-hop Eclipse iteratively chooses a sequence of configurations that maximizes the cost-adjusted
utility, just like Eclipse, but the cost-unadjusted utility $U(M, \alpha)$ here accounts for not only the traffic that can be transmitted under direct routing, but also that can be indirectly routed over all possible 2-hop paths.


\subsection{A Qualified Analogy}\label{sec:intuition}

We make a qualified analogy between this scheduling of the circuit switch and the scheduling of ``flights''.  We view the connections
(between the input ports and the output ports) in a matching $M_k$ as ``disjoint flights'' (those that share neither a source nor a destination ``airport'') and
the residue capacity on such a connection as ``available seats''.  We view Eclipse, Eclipse++, and 2-hop Eclipse as different ``flight booking''
algorithms.  Eclipse books ``passengers''
 (traffic in the demand matrix) for ``non-stop flights'' only.  Then Eclipse++ books the ``remaining
passengers'' for ``flights with stops'' using only the ``available seats'' left over after Eclipse does its ``bookings''.  Different than Eclipse++, 2-hop Eclipse
``books passengers'' for both ``non-stop'' and ``one-stop flights'' early on, although it does try to put ``passengers'' on ``non-stop
flights'' as much as possible, since each ``passenger'' on a ``one-stop flight'' costs twice as many ``total seats'' as that on a
``non-stop flight''.

It will become clear shortly that the sole purpose of this qualified analogy is for us to distinguish two types of ``passengers'' in presenting
the 2-hop Eclipse algorithm:
those ``looking for a non-stop flight'' whose counts are encoded as the remaining demand matrix $D_{\mbox{\footnotesize rem}}$,
and those ``looking for a connection flight'' to complete their potential ``one-stop itineraries'', whose counts are
encoded as a new $n\times n$ matrix $I_{\mbox{\footnotesize rem}}$ that we will describe shortly.  We emphasize that
this analogy shall not be stretched any further, since it would otherwise lead to absurd inferences, such as that such a set of disjoint ``flights''
must span the same time duration and have the same number of ``seats'' on them.

\subsection{The Pseudocode}

The pseudocode of 2-hop Eclipse is shown in Algorithm~\ref{alg:Eclipse}.
It is almost identical to that of Eclipse \cite{bojjacostly}.  The only major difference is that in each iteration (of the ``while'' loop), 2-hop Eclipse searches for a matching
$(M, \alpha)$ that maximizes $\frac{\|\min(\alpha M,D_{\mbox{\footnotesize rem}}+I_{\mbox{\footnotesize rem}})\|_1}{\delta+\alpha}$, whereas Eclipse searches for one that maximizes $\frac{\|\min(\alpha M,D_{\mbox{\footnotesize rem}})\|_1}{\delta+\alpha}$. In other words, in each iteration, 2-hop Eclipse first performs some preprocessing to obtain $I_{\mbox{\footnotesize rem}}$ and then substitute the parameter $D_{\mbox{\footnotesize rem}}$ by
$D_{\mbox{\footnotesize rem}} + I_{\mbox{\footnotesize rem}}$ in making the ``argmax'' call (Line 7).   The ``while'' loop of Algorithm \ref{alg:Eclipse} terminates when every row or column sum of $D_{\mbox{\footnotesize rem}}$ is no more than $r_p t_c$, where $r_p$ denotes the (per-port) transmission rate of the packet switch and $t_c$ denotes the total transmission time used so far by the circuit switch, since the remaining
traffic demand can be transmitted by the packet switch (in $t_c$ time).   Note there is no occurrence of $r_c$, the (per-port) transmission rate of the circuit switch, in Algorithm \ref{alg:Eclipse},
because we normalize $r_c$ to $1$ here and throughout this paper.

\begin{algorithm}[t]
\caption{2-hop Eclipse}
\label{alg:Eclipse}
    \KwIn {Traffic demand $D$;}
    \KwOut {Sequence of schedules $(M_k, \alpha_k)_{k=1,...,K}$;}
    sch $\leftarrow$ \{\};\Comment{schedule}\\
    $D_{\mbox{\footnotesize rem}}$ $\leftarrow$ $D$;\Comment{remaining demand}\\
    $R$ $\leftarrow$ \textbf{0};\qquad\qquad \Comment{residue capacity}\\
    $t_c$ $\leftarrow$ 0;\qquad\qquad \Comment{transmission time of circuit switch}\\
    \While{$\exists$ any row or column sum of $D_{\mbox{\footnotesize rem}} > r_p t_c$ }
        {
                Construct $ I_{\mbox{\footnotesize rem}} $ from $(D_{\mbox{\footnotesize rem}},R)$;\quad \Comment{2-hop demand matrix}\label{step:demand_extension}\\
                $(M, \alpha)\leftarrow\mathop{\arg\max}\limits_{M\in\mathcal{M},\alpha\in \mathbb{R}_+} \frac{\|\min(\alpha M,D_{\mbox{\footnotesize rem}}+I_{\mbox{\footnotesize rem}})\|_1}{\delta+\alpha}$;\label{step:greedymax}\\
                sch $\leftarrow$ sch $\cup\, \{(M,\alpha)\}$;\\
                $t_c \leftarrow t_c+\delta+\alpha $;\\
                Update $D_{\mbox{\footnotesize rem}}$;\label{step:updateD}\\
                Update $R$;\label{step:updateR}
        }
\end{algorithm}


\subsection{The Matrix \texorpdfstring{$I_{\mbox{\footnotesize rem}}$}{Lg}}

Just like $D_{\mbox{\footnotesize rem}}$, the value of $I_{\mbox{\footnotesize rem}}$ changes after each iteration.
We now explain the value of $I_{\mbox{\footnotesize rem}}$, at the beginning of the $k^{th}$ iteration ($k > 1$).
To do so, we need to first introduce another matrix $R$.
As explained earlier, among the edges that belong to the matchings $(M_1, \alpha_1)$, $(M_2, \alpha_2)$, $\cdots$, $(M_{k-1}, \alpha_{k-1})$ computed
in the previous $k-1$ iterations, some may have residue capacities.   These residue capacities are captured in an $n\times n$ matrix $R$ as follows:
$R(l, i)$ is the total residue capacity of all edges from input $l$ to output $i$ that belong to one of these (previous) $k-1$ matchings.  Under the qualified analogy above,
$R(l, i)$ is the total number of ``available seats on all previous flights from airport $l$ to airport $i$''.   We refer to $R$ as the \textit{(cumulative) residue capacity matrix}
in the sequel.

Now we are ready to define $I_{\mbox{\footnotesize rem}}$.  Consider that, at the beginning of the $k^{th}$ iteration, $D_{\mbox{\footnotesize rem}}(l,j)$ ``local passengers''
(\textit{i.e.}, those who are originated at $l$) who need to fly to $j$ remain to have their ``flights'' booked.  Under Eclipse, they have to be booked on either a ``non-stop flight'' or a ``bus''
(\textit{i.e.}, through the packet switch) to $j$.  Under 2-hop Eclipse, however, there is a third option:  a ``one-stop flight'' through
an intermediate ``airport''.  2-hop Eclipse explores this option as follows.  For each possible intermediate ``airport'' $i$ such that $R(l, i) > 0$ (\textit{i.e.}, there are ``available seats''
on one or more earlier ``flights'' from $l$ to $i$), $I^{(l)}_{\mbox{\footnotesize rem}}(i,j)$ ``passengers'' will be on the ``speculative standby list'' at ``airport'' $i$, where
\begin{equation}\label{eq:shiftable_demand}
I^{(l)}_{\mbox{\footnotesize rem}}(i,j)\triangleq\min\big{(}D_{\mbox{\footnotesize rem}}(l,j),R(l,i)\big{)}.
\end{equation}
In other words, up to $I^{(l)}_{\mbox{\footnotesize rem}}(i,j)$ ``passengers'' could be booked on ``earlier flights'' from $l$ to $i$ that have a total of $R(l,i)$ ``available seats'' on them,
and ``speculatively stand by'' for a ``flight'' from $i$ to $j$ that might materialize as a part of the matching $M_k$.

The matrix element $I_{\mbox{\footnotesize rem}}(i,j)$ is the total number of ``nonlocal passengers'' who are originated at all ``airports'' other than $i$ and $j$
and are on the ``speculative standby list'' for a possible ``flight'' from $i$ to $j$.  In other words, we have
\begin{equation}\label{eq:demand_extension}
I_{\mbox{\footnotesize rem}}(i,j)\triangleq\sum_{l\in [n]\setminus\{i,j\}}I^{(l)}_{\mbox{\footnotesize rem}}(i,j).
\end{equation}
Recall that $D_{\mbox{\footnotesize rem}}(i,j)$ is the number of ``local passengers'' (at $i$) that need to travel to $j$.
Hence at the ``airport'' $i$, a total of $D_{\mbox{\footnotesize rem}}(i,j) + I_{\mbox{\footnotesize rem}}(i,j)$ ``passengers'', ``local or nonlocal'', could
use a ``flight'' from $i$ to $j$ (if it materializes in $M_k$).
We are now ready to precisely state the different between Eclipse and 2-hop Eclipse:
Whereas $\|\min(\alpha M, D_{\mbox{\footnotesize rem}})\|_1$, the cost-unadjusted utility function used by Eclipse, accounts only for
``local passengers'', $\|\min(\alpha M, D_{\mbox{\footnotesize rem}}+I_{\mbox{\footnotesize rem}})\|_1$, that used by 2-hop Eclipse, accounts for both
``local'' and ``nonlocal passengers''.

Note that the term $D_{\mbox{\footnotesize rem}}(l,j)$ appears in the definition of $I^{(l)}_{\mbox{\footnotesize rem}}(i,j)$ (Formula (\ref{eq:shiftable_demand})),
for all $i\in[n]\setminus \{l,j\}$.
In other words, ``passengers'' originated at $l$ who need to travel to $j$ could be on the ``speculative standby list'' at multiple intermediate ``airports''.   This is however
not a problem
(\textit{i.e.}, will not result in ``duplicate bookings'') because at most one of these ``flights'' (to $j$) can materialize as a part of matching $M_k$.

\subsection{Update \texorpdfstring{$D_{\mbox{\footnotesize rem}}$}{Lg} and \texorpdfstring{$R$}{Lg}}

After the schedule $(M_k, \alpha_k)$ is determined by the ``argmax call'' (Line~\ref{step:greedymax} in Algorithm~\ref{alg:Eclipse}) in the $k^{th}$ iteration,
the action should be taken on ``booking'' the right set of ``passengers'' on the ``flights'' in $M_k$, and updating
$D_{\mbox{\footnotesize rem}}$ (Line~\ref{step:updateD}) and $R$ (Line~\ref{step:updateR}) accordingly.
Recall that we normalize $r_c$, the service rate of the circuit switch, to $1$, so all these flights have $\alpha_k \times 1 = \alpha_k$ ``available seats''.
We only describe how to do so for a single ``flight'' (say from $i$ to $j$) in $M_k$; that for other ``flights'' in $M_k$ is similar.
Recall that $D_{\mbox{\footnotesize rem}}(i,j)$ ``local passengers'' and $I_{\mbox{\footnotesize rem}}(i,j)$ ``nonlocal passengers''
are eligible for a ``seat'' on this ``flight''.
When there are not enough seats for all of them, 2-hop Eclipse prioritizes ``local passengers'' over ``nonlocal passengers'', because the former
is more resource-efficient to serve than the latter, as explained earlier.
There are three possible cases to consider:

\begin{spenumerate}
\item \emph{$\alpha \leq D_{\mbox{\footnotesize rem}}(i,j)$}.  In this case, only a subset of ``local passengers'' (directly routed traffic), in the ``amount'' of
$\alpha_k$, are booked on
this ``flight'', and $D_{\mbox{\footnotesize rem}}(i,j)$ is hence decreased by $\alpha$.
There is no ``available seat'' on this ``flight'' so the value of $R(i, j)$ is unchanged.

\item \emph{$\alpha \geq D_{\mbox{\footnotesize rem}}(i,j)+I_{\mbox{\footnotesize rem}}(i,j)$}. In this case, all ``local'' and ``nonlocal passengers''
are booked on this ``flight''.  Note that, for any $l\in [n]\setminus\{i,j\}$,
$I^{(l)}_{\mbox{\footnotesize rem}}(i,j)$ among the $I_{\mbox{\footnotesize rem}}(i,j)$ ``nonlocal passengers'',
are originated at $l$.  These $I^{(l)}_{\mbox{\footnotesize rem}}(i,j)$ ``nonlocal passengers'' are booked on one or more ``earlier flights''
from $l$ to $i$, and also on this ``flight'' to complete their 2-hop ``itineraries''.   After all these ``bookings'',
$D_{\mbox{\footnotesize rem}}(i,j)$ is set to $0$ (all ``local passengers'' traveling to $j$ gone), and for each $l\in [n]\setminus\{i,j\}$,
$D_{\mbox{\footnotesize rem}}(l,j)$ and $R(l,i)$ each is decreased by $I^{(l)}_{\mbox{\footnotesize rem}}(i,j)$ to account for the
resources consumed by the indirect routing of traffic demand (\textit{i.e.}, ``nonlocal passengers''), in the amount of $I^{(l)}_{\mbox{\footnotesize rem}}(i,j)$,
from $l$ to $j$ via $i$.  Finally, $R(i,j)$ is increased by $\alpha-\big{(}D_{\mbox{\footnotesize rem}}(i,j)+I_{\mbox{\footnotesize rem}}(i,j)\big{)}$,
the number of ``available seats'' that remain on this flight after all these ``bookings''.


\item \emph{$D_{\mbox{\footnotesize rem}}(i,j)<\alpha<D_{\mbox{\footnotesize rem}}(i,j)+I_{\mbox{\footnotesize rem}}(i,j)$}.
In this case, all ``local passengers''
are booked on this ``flight'', so $D_{\mbox{\footnotesize rem}}(i,j)$ is set to $0$.
However, different from the previous case, there are not enough ``available seats'' left on this ``flight'' to accommodate all
$I_{\mbox{\footnotesize rem}}(i,j)$ ``nonlocal passengers'',
so only a proper ``subset'' of them
can be booked on this ``flight''.  We allocate this proper ``subset'' proportionally to all origins $l\in [n]\setminus\{i,j\}$.
More specifically, for each $l\in [n]\setminus\{i,j\}$, we book $\theta\cdot I^{(l)}_{\mbox{\footnotesize rem}}(i,j)$ ``nonlocal passengers''
originated at $l$ on one or more ``earlier flights''
from $l$ to $i$, and also on this ``flight'', where $\theta\triangleq\frac{\alpha-D_{\mbox{\footnotesize rem}(i,j)}}{I_{\mbox{\footnotesize rem}}(i,j)}$.
Similar to that in the previous case, after these ``bookings'', $D_{\mbox{\footnotesize rem}}(l,j)$ and $R(l,i)$ each is decreased by $\theta\cdot I^{(l)}_{\mbox{\footnotesize rem}}(i,j)$. Finally, $R(i,j)$ is unchanged as this ``flight'' is full.


\end{spenumerate}


It remains to discuss a subtle issue in the above actions of ``booking passengers'' and updating the ``seat maps''.
In case (3) above, we allocate the ``available seats'' to ``nonlocal passengers'' in a proportional manner.
We have tried other allocation techniques, such as water-filling, VOQ with the the largest residue capacity first
(\textit{i.e.}, $\mathop{\arg\max}_{l\in [n]\setminus\{i,j\}} D(l,j)$), and VOQ with the smallest residue capacity first, but none of them
resulted in better performance.  This outcome is not surprising because the proportional allocation is
quite efficient in reducing the sizes of these ``leftover'' (after being served by ``non-stop flights'') VOQs.  When
they all become small in sizes eventually, the packet switch can take care of all of them, even though many of them
remain to have nonzero sizes.

\subsection{Complexity of 2-hop Eclipse}\label{sec:complexity_twohop}

Each iteration in 2-hop Eclipse has only a slightly higher computational complexity than that in Eclipse.  This additional complexity
comes from Lines~\ref{step:demand_extension},~\ref{step:updateD}, and~\ref{step:updateR} in Algorithm~\ref{alg:Eclipse}.
We need only to analyze the complexity of line~\ref{step:demand_extension} (for updating $I^{(l)}_{\mbox{\footnotesize rem}}$), since it
dominates those of others.  For each $k$, the complexity of line~\ref{step:demand_extension} in the $k^{th}$ iteration is $O(kn^2)$ because there were
at most $(k-1)n$ ``flights'' in the past $k-1$ iterations, and for each such flight (say from $l$ to $i$), we need to update at most $n-2$ variables, namely
$I^{(l)}_{\mbox{\footnotesize rem}}(i,j)$ for all $j\in [n]\setminus\{l,i\}$.  Hence the total additional complexity across all iterations is $O(\min(K,n)Kn^2)$, where $K$ is the
number of iterations actually executed by 2-hop Eclipse.  Adding this to $O(n^{5/2}\log n\log B)$, the complexity of Eclipse, we arrive at the complexity of 2-hop Eclipse:
$O(Kn^{5/2}\log n\log B+\min(K,n)Kn^2)$ (see \autoref{tab:complexity}). Finally, as shown in~\autoref{tab:complexity}, the complexity of Eclipse++ is much higher than those of both Eclipse and 2-hop Eclipse, as are confirmed by our experiments. Here $W$ denotes the maximum row/column sum of the demand matrix.


\begin{table}[h]
  \centering
    \begin{tabular}{c|l}
    \hline
    \hline
    \textbf{Algorithm} &  \textbf{Time Complexity}  \\
    \hline
    Eclipse & $O(Kn^{5/2}\log n\log B)$\\ \hline
    2-hop Eclipse & $O(Kn^{5/2}\log n\log B+\min(K,n)Kn^2)$\\ \hline
    Eclipse++ & $O(WKn^3(\log K+\log n)^2)$\\ \hline\hline 
    \end{tabular}%
    \caption{Comparison of time complexities}
  \label{tab:complexity}%
\end{table}


\subsection{Parallel Eclipse and 2-hop Eclipse}\label{sec:parallel_eclipse}


As we will show shortly in \autoref{sec:execution_time}, the execution time of 2-hop Eclipse is quite long (about one second for $n = 32$ and tens of seconds for $n = 100$ using a desktop 
computer) and that of Eclipse is only a bit shorter.   This motivates us to speedup both Eclipse and 2-hop Eclipse via parallelization.  Since the only computationally heavy step 2-hop Eclipse has on top of Eclipse's is Line~\ref{step:demand_extension} in Algorithm~\ref{alg:Eclipse}, which can be straightforwardly parallelized (without resorting to approximate computation) 
due to the lack of data dependency therein, we describe mostly how Eclipse is parallelized in the sequel.  


In each ``argmax'' call (Line~\ref{step:greedymax}) in Algorithm~\ref{alg:Eclipse}, 
Eclipse has to perform roughly $\log_2(n^2) \times 2 = 4 \log_2 n$ Maximum Weight Matching (MWM) computations 
to binary-search for the (empirically) optimal value of the parameter 
$\alpha_k$ over the space of $n^2$ possible values between $0$ and $1$ (the $n^2$ elements of the matrix $D_{\mbox{\footnotesize rem}}$), where each search step involves $2$ MWM computations.  Since a MWM computation is very expensive, this ``argmax'' call accounts for the bulk of the computation time of (each iteration of) Algorithm~\ref{alg:Eclipse}.  
Hence, ideally we would like to parallelize these $4 \log_2 n$ MWM computations.  Unfortunately, they cannot be easily parallelized since the next MWM computation to perform (\textit{i.e.}, the next parameter value to try) depends on the outcome of the current MWM computation. While trying all $n^2$ parameter values in parallel can do the trick, $n^2$ processors have to be used, which is extremely expensive when $n$ is large.

Our solution is to strategically sample a much smaller (than $n^2$) subset of possible parameter values and search in parallel only for the optimal parameter value within this subset.
Hence, there is approximate computation involved in this parallelization step.  However, the scheduling performance loss due to this approximation is very small:  
As will be shown in \autoref{sec:eval_parallel_2hop_eclipse}, even when the 2-hop Eclipse searches over a subset of only $O(n)$ distinct values (\textit{i.e.}, requires only $O(n)$ parallel processors) sampled naively, the transmission time performance of the resulting algorithm is either similar to or slightly worse than that of the un-sampled algorithm (that searches over the set of all $n^2$ values).   This parallelization would shorten the execution time of Eclipse, and that of 2-hop Eclipse, by roughly $4 \log_2 n$ (=20 when $n = 32$) times.  Hence for a small data center with $n = 32$ racks, each parallel 2-hop Eclipse computation can finish within tens of milliseconds, 
which is fast enough for the type of latency typically expected of hybrid-switched traffic\footnote{A small amount of real-time mission-critical traffic can be fast-tracked through the packet switch without waiting for this 
computation.}.

\subsection{Why 2-hop Only?}\label{sec:constraint}

We restrict indirect routing to most 2 hops in 2-hop Eclipse, due to the following three reasons.  First, it can be shown that the aforementioned ``duplicate bookings'' could happen
if indirect routing of $3$ or more hops are allowed.  Second, the optimization problem of maximizing the cost-adjusted utility can no longer be precisely
formulated as maximum-weighted matching (MWM) computation and hence
becomes much harder to solve, if not NP-hard.   For one thing, the residue capacity matrix $R$, which is used to calculate the indirect traffic demand matrix
$I_{\mbox{\footnotesize rem}}$, does not capture the chronological order between any two ``flights'' with ``available seats'';  whereas this ordering information
is not needed in 2-hop indirect routing, it is when 3 or more hops is allowed.
Last but not the least, 2-hop indirect routing appears to have reaped most of the performance benefits indirect routing could possibly
afford us, as shown in \autoref{sec:evaluation_eclipseplus}.  This intuitively makes sense:  Since each extra hop consumes the bandwidth resource of an extra link, indirect routing quickly becomes very resource-inefficient when the
number of hops is larger than 2.

\section{Scheduling Algorithm in Partially reconfigurable circuit switch}\label{sec:partial_reconfig2}

In this section, we allow the circuit switch to become partially reconfigurable in the sense that
only the input ports affected by the configuration change need to pay a reconfiguration delay $\delta$.
We propose a low-complexity scheduling algorithm called BFF (best first fit) that can take full advantage of this capability to further improve performance.
BFF is the first algorithmic solution that exploits partial reconfiguration in hybrid switching.


\subsection{Problem Statement}\label{sec:partial_reconfig}

The objective of this scheduling remains the same as before:
given a demand matrix $D$, to find a \textit{schedule} for the circuit switch that, to the extent possible,
minimizes the total transmission time.  However, with this partial reconfiguration capability,
matchings between input ports and output ports no longer need to start and stop in synchrony.   Hence the
notations $(M_1, \alpha_1)$, $(M_2, \alpha_2)$, $\cdots$, $(M_K, \alpha_K)$ are no longer suitable for representing the circuit switch schedule.
Instead, we use an $n\times n$ matrix process $S(t)=\big{(}s_{ij}(t)\big{)}$ for that.
For any given time $t$, $S(t)$ is a $0-1$ (sub-matching) matrix that encodes the matching between input ports and
the output ports at time $t$.  More specifically, $s_{ij}(t) = 1$ if input port $i$ is connected to output $j$ at time $t$,
and $s_{ij}(t) = 0$ otherwise.  Note that, in this section, we are using the standard notation for the matrix element $s_{ij}$, which is different than that used in earlier sections such as $R(i,j)$.





\subsection{Background on Open-Shop Scheduling}\label{sec:intuition_bff}

When partial reconfiguration is allowed, there is no packet switch, and only direct routing is considered,
scheduling the circuit switch alone can be converted into the classical \textit{open-shop scheduling problem} (OSS)~\cite{pinedo2016scheduling,towles2003guaranteed,van2017adaptive}.
In an OSS problem, there are a set of $N$ jobs,
a set of $m$ machines, and a two-dimensional table specifying the amount of time (could be $0$)
that a job must spend at a machine to have a certain task performed. The scheduler has to assign jobs to machines in such a way, that at any moment of time,
no more than one job is assigned to a machine and no job is assigned to more than one machine\footnote{In this respect, OSS is different than 
\textit{concurrent OSS}~\cite{mastrolilli2010minimizing}, which allows any job to be concurrently processed by multiple machines at the same time.}. The mission is accomplished when every job has all its tasks
performed at respective machines.  The OSS problem is to design an algorithm that minimizes, to the extent possible,
the makespan of the schedule, or the amount of time it takes to accomplish the mission.
In this circuit switching (only) problem, input ports are jobs, output ports are machines,
each VOQ$(i, j)$ is a task that belongs to job $i$ and needs to be performed at machine $j$ for the amount of time $D(i, j)$.
In OSS, a machine may need some time to reconfigure between taking on a new job, which corresponds to the reconfiguration delay $\delta$
in circuit switching. The OSS problem is in general NP-hard, so only heuristic or approximate solutions to it~\cite{shmoys1994improved,hochba1997approximation,brasel2008heuristic} exist that run in polynomial time.


There are two types of OSS problems:  one that allows a task to be preempted at a machine for another task (\textit{i.e.}, preemptive) and one that does not
(\textit{i.e.}, non-preemptive).   In this circuit switch (only) scheduling problem, preemption means that the traffic in a VOQ$(i, j)$ is split into multiple bursts to be
sent over multiple noncontiguous connections between $i$ and $j$.   For this circuit switching (only) problem, non-preemptive solutions generally do not perform
worse because each preemption costs us a nontrivial reconfiguration delay $\delta$, and such preemption costs can hardly be compensated by the
larger solution space and more scheduling flexibility that preemptive solutions can provide.

\subsection{LIST: A Family of Heuristics}

LIST (list scheduling) is a well-known family of polynomial-time heuristic OSS
algorithms~\cite{shmoys1994improved,hochba1997approximation,brasel2008heuristic}.
LIST starts by attempting to assign an available job (\textit{i.e.}, not already being worked on by a machine)
to one of the available machines on which the job has a task to perform, according to a machine preference criterion
(can be job-specific and time-varying).
If multiple jobs are competing for the same machine, one of the jobs is chosen according to a job preference criterion
(can be machine-specific and time-varying).
After all initial assignments are made, the scheduler ``sits idle'' until a task is completed on a machine, in which
case both the corresponding job and the machine become available.
Once a machine becomes available, any available job that has a task to be performed on the machine can compete for the machine.

Our BFF algorithm is a slight adaptation of a non-preemptive LIST algorithm \cite{brasel2008heuristic} that uses
LPT (longest processing time) as the preference criterion for both the machines and the jobs.
In LPT LIST, whenever multiple jobs
compete for a machine, the machine picks the ``most time-consuming task'', \textit{i.e.}, the job that
takes the longest time to finish on the machine;  whenever a job has multiple available machines to choose from, it chooses among them the machine that has
the ``most time-consuming task'' to perform on the job.  In other words, LPT gives preference to longer tasks, whether a machine is choosing
jobs or a job is choosing machines.  LPT is a perfect match for our problem, because with a packet switch to ``sweep clean'' all
short tasks (\textit{i.e.}, tiny amounts of remaining traffic left over in VOQs by the circuit switch), the circuit switch can afford to focus only on a comparatively
small number of long tasks.

However, no algorithm in the LIST family, including LPT LIST, is a good fit for the problem of circuit switching only (\textit{i.e.}, no packet switch), when the circuit switch is partially reconfigurable~\cite{towles2003guaranteed}.
In particular, it was shown in~\cite{towles2003guaranteed} that, whenever a scheduling algorithm from the LIST family is used, whether the circuit switch is partially reconfigurable or not
makes almost no difference in the the performance (measured by transmission time) of the resulting schedule.
This is because, without the help from a packet switch,
the circuit switch would have to ``sweep clean'' the large number of short tasks (VOQs) all by itself, and each such short task costs the circuit switch
a reconfiguration delay $\delta$ that is significant compared to its processing (transmission) time.   To the best of our knowledge, BFF is the first time
that a LIST algorithm is adapted for hybrid switching.

\subsection{BFF Algorithm}

As explained earlier, BFF is a slight adaptation of the LPT LIST algorithm for open-shop scheduling.  There are two (minor) differences between BFF and
LPT LIST.  First, at the beginning of the scheduling (\textit{i.e.}, $t = 0$), when all jobs and all machines are available, BFF runs the aforementioned
$O(n^{5/2} \log B)$ maximum weighted matching (MWM) algorithm~\cite{duan2012scaling} to obtain the heaviest (\textit{w.r.t.} to their weights $D$) initial matching between jobs and machines.
BFF does not use LPT LIST for this initialization step because it would likely result in a sub-optimal (\textit{i.e.}, lighter in weight) matching to start with.
Second, like 2-hop Eclipse, BFF terminates when the remaining demand $D_{\mbox{\footnotesize rem}}$ becomes small enough for the packet switch to handle.

\begin{algorithm}[t]
\caption{Action taken by BFF after a machine is done with a job.}
\label{alg:BFF}
  \vspace{0.5em}
    \SetKwProg{Proc}{When input port $i$ finishes transmitting VOQ$(i, j)$ to output port $j$ at time $\tau$:}{}{}

    \nonl \Proc{}{
            Output\_Seek\_Pairing$(j,\tau)$;\\
            \label{step:outputpairing}

            Input port $i$ reconfigures during $[\tau,\tau+\delta]$;\\
            \label{step:start_reconfig}

            Input\_Seek\_Pairing$(i,\tau+\delta)$;\\
            \label{step:inputpairing}
    }
    \vspace{0.5em}

    \SetKwProg{Proce}{Procedure}{}{}

    \nonl \Proce{Output\_Seek\_Pairing$(j,t)$}{
        Update $D_{\mbox{\footnotesize rem}}$;\label{step:update2}\\
        \uIf {$\{l\in I_a\mid D_{\mbox{\footnotesize rem}}(l,j)>0\}\neq\phi$}{
            $l=\mathop{\arg\max}_u D_{\mbox{\footnotesize rem}}(u,j)$;\\
            Connect output $j$ with input $l$;\label{step_connect1}
        }\Else{$O_a\leftarrow O_a\cup\{j\}$;}
    }\label{step:end1}
    \vspace{0.5em}


    \nonl \Proce{Input\_Seek\_Pairing$(i,t)$}{
        Update $D_{\mbox{\footnotesize rem}}$;\label{step:update3}\\
        \uIf {$\{j\in O_a\mid D_{\mbox{\footnotesize rem}}(i,j)>0\}\neq\phi$}{
            $j=\mathop{\arg\max}_v D_{\mbox{\footnotesize rem}}(i,v)$;\\
            Connect input $i$ with output $j$;\label{step_connect2}
        }\Else{$I_a\leftarrow I_a\cup\{i\}$;}
    }\label{step:end2}
\end{algorithm}


In BFF, for each input port $i_1$, the task of deciding with which outputs the input port $i_1$ should be matched with over time
is almost independent of that for any other input port $i_2$.   Hence, to describe BFF precisely, it
suffices to describe the actions taken by the scheduler after a job $i$ get its task performed
at a machine $j$ (\textit{i.e.}, after input port $i$ transmits all traffic in VOQ$(i, j)$, in the amount of $D(i, j)$, to output port $j$).

We do so in Algorithm~\ref{alg:BFF}. Suppose the machine $j$ is done with the job $i$ at time $\tau$.
Then machine (output port) $j$ immediately looks to serve another job by calling ``Output\_Seek\_Pairing$(j,t)$''
(Line \ref{step:outputpairing} in Algorithm~\ref{alg:BFF}).
The job (input port) $i$, on the hand, is not ready to be performed on another machine (output port) until
$\tau + \delta$ (\textit{i.e.}, after a reconfiguration delay), so it calls ``Input\_Seek\_Pairing$(i,\tau+\delta)$''
(Line \ref{step:inputpairing} in Algorithm~\ref{alg:BFF}).   However, since in this batching scheduling setting,
the whole schedule $S(t)$ is computed before any transmission (according to the schedule) can begin,
input port $i$ knows which machine it will be paired with at time $\tau + \delta$.  Hence input port $i$ can
start reconfiguring to pair with that machine at time $\tau$ (Line \ref{step:inputpairing} in Algorithm~\ref{alg:BFF})
so that the actual transmission can start at time $\tau + \delta$.

In Algorithm~\ref{alg:BFF}, $I_a$ and $O_a$ denote the sets of available jobs (input ports) and of available machines (output ports) respectively.
Clearly, Procedure ``Output\_Seek\_Pairing( )'' (Lines \ref{step:update2} through \ref{step:end1}) follows the
LPT (longest processing time first) preference criteria:  It tries to identify, for the machine (output port) $j$,
the job (input port) that brings with it the largest task, among the set of available jobs $I_a$.  Similar things
can be said about procedure ``Input\_Seek\_Pairing( )'' (Lines \ref{step:update3} through \ref{step:end2}).
In other words, each output port, at the very {\it first} moment
it becomes available (to input ports), attempts to match with the {\it best} input port (\textit{i.e.}, the one with the largest amount of work
for it to do), and vice versa.  Therefore, we call our algorithm BFF (Best First Fit).

In addition to the $O(n^{5/2}\log B)$ complexity needed to obtain an
MWM (using~\cite{duan2012scaling}) at the very beginning, with a straightforward implementing using a straightforward data structure,
BFF has a computational complexity
of $O(Kn^2)$, where $K$ is the average number of times each input port needs to reconfigure over time.
Here we reused and arguably slightly abused the notation $K$ used in \autoref{sec:indirect_routing} to denote the total
number of matchings over time.  Hence the overall complexity of BFF is $O(Kn^2+n^{5/2}\log B)$, which is asymptotically
smaller than that of Eclipse, which is $O(Kn^{5/2}\log n\log B)$ (see Table \ref{tab:complexity}).   Empirically, BFF runs about three orders of magnitude faster than
Eclipse, as will be shown in~\autoref{sec:execution_time}.

\subsection{Combine BFF with Indirect Routing}

As mentioned earlier, BFF considers only direct routing.  Readers might wonder if allowing indirect routing can bring further performance improvements.
However, after several attempts at combining indirect routing with BFF, we realize that this direction is a dead end for two reasons.
First, BFF leaves
little ``slack'' in the schedule for the indirect routing to gainfully exploit.  Second, any indirect routing, even using the relatively computationally efficient
2-hop Eclipse algorithm, increases the computational complexity of BFF significantly.

\section{Evaluation}\label{sec:evaluation}
In this section, we evaluate the performances of our solutions 2-hop Eclipse and BFF, and compare them with those
of Eclipse and Eclipse++, under various system parameter settings and traffic demands.
We do not however have Eclipse++ in all performance figures because its computational complexity is so high that it usually takes a few hours to
compute a schedule.  However, those Eclipse++ simulation results we managed to obtain and present in \autoref{sec:evaluation_eclipseplus} show conclusively that
the small reductions in transmission time using Eclipse++ are not worth its extremely high computational complexity.
We do not compare our solutions with Solstice~\cite{liu2015scheduling} in these evaluations, since Solstice was shown in~\cite{bojjacostly} to perform worse than Eclipse in
all simulation scenarios.  For all these comparisons,
we use the same performance metric as that used in~\cite{liu2015scheduling}:  the total time needed for
the hybrid switch to transmit the traffic demand $D$.








\subsection{Traffic Demand Matrix \texorpdfstring{$D$}{Lg}}

For our simulations, we use the same traffic demand matrix $D$ as used in other hybrid scheduling works~\cite{liu2015scheduling,bojjacostly}.
In this matrix, each row (or column)
contains $n_L$ large equal-valued elements (large input-output flows) that as a whole account for $c_L$ (percentage) of the total workload to the row (or column),
$n_S$ medium equal-valued elements (medium input-output flows) that as a whole account for the rest $c_S = 1- c_L$ (percentage), and noises.
Roughly speaking, we have
\begin{equation}\label{eq:traffic_matrix}
D=(\sum\limits_{i=1}^{n_L} \frac{c_L}{n_L}P_i+\sum\limits_{i=1}^{n_S} \frac{c_S}{n_S}P'_{i}+\mathcal{N}_1)\times 90\%+\mathcal{N}_2
\end{equation}
where $P_i$ and $P'_{i}$ are random $n\times n$ matching (permutation) matrices.

The parameters $c_L$ and $c_S$ control the aforementioned skewness (few large elements in a row or column account for the majority of the row or column
sum) of the traffic demand.
Like in~\cite{liu2015scheduling,bojjacostly}, the default values of $c_L$ and $c_S$ are
$0.7$ (\textit{i.e.}, $70\%$) and $0.3$ (\textit{i.e.}, $30\%$) respectively, and the default values of $n_L$ and $n_S$ are $4$ and $12$ respectively.
In other words, in each row (or column) of the demand matrix, by default the $4$ large flows
account for $70\%$ of the total traffic in the row (or column), and the $12$ medium flows account for the rest $30\%$.   We will also study how these hybrid
switching algorithms perform when the traffic demand has other degrees of skewness by varying $c_L$ and $c_S$.



As shown in Equation~(\ref{eq:traffic_matrix}),
we also add two noise matrix terms $\mathcal{N}_1$ and $\mathcal{N}_2$ to $D$.
Each nonzero element in $\mathcal{N}_1$ is a Gaussian random variable that is to be added to a traffic demand matrix element that was nonzero
before the noises are added.  This noise matrix $\mathcal{N}_1$ was also used in~\cite{liu2015scheduling,bojjacostly}.
However, each nonzero (noise) element here in $\mathcal{N}_1$ has a larger standard deviation, which is equal to $1/5$ of the value of
the demand matrix element it is to be added to, than that in \cite{liu2015scheduling,bojjacostly}, which is equal to
$0.3\%$ of $1$ (the normalized workload an input port receives during a scheduling window, \textit{i.e.}, the sum of the corresponding row in $D$). We increase this additive noise here to highlight the performance robustness of our algorithm to such perturbations.

Different than in~\cite{liu2015scheduling,bojjacostly}, we also add (truncated) positive Gaussian noises $\mathcal{N}_2$ to a portion of the zero entries in the demand matrix
in accordance with the following observation.
Previous measurement studies have shown that ``mice flows'' in the demand matrix are heavy-tailed~\cite{benson2010network}
in the sense the total traffic volume of these ``mice flows'' is not insignificant.
To incorporate this heavy-tail behavior (of ``mice flows'') in the traffic demand matrix, we add such a positive Gaussian noise -- with standard deviation equal to $0.3\%$ of
$1$ -- to $50\%$ of the zero entries of the demand matrix. This way the ``mice flows'' collectively
carry approximately $10\%$ of the total traffic volume.  To bring the normalized workload back to $1$, we scale the demand matrix by $90\%$ before adding
$\mathcal{N}_2$, as shown in (\ref{eq:traffic_matrix}).



\begin{figure*}[b]
\centering
\includegraphics[width=0.95\textwidth]{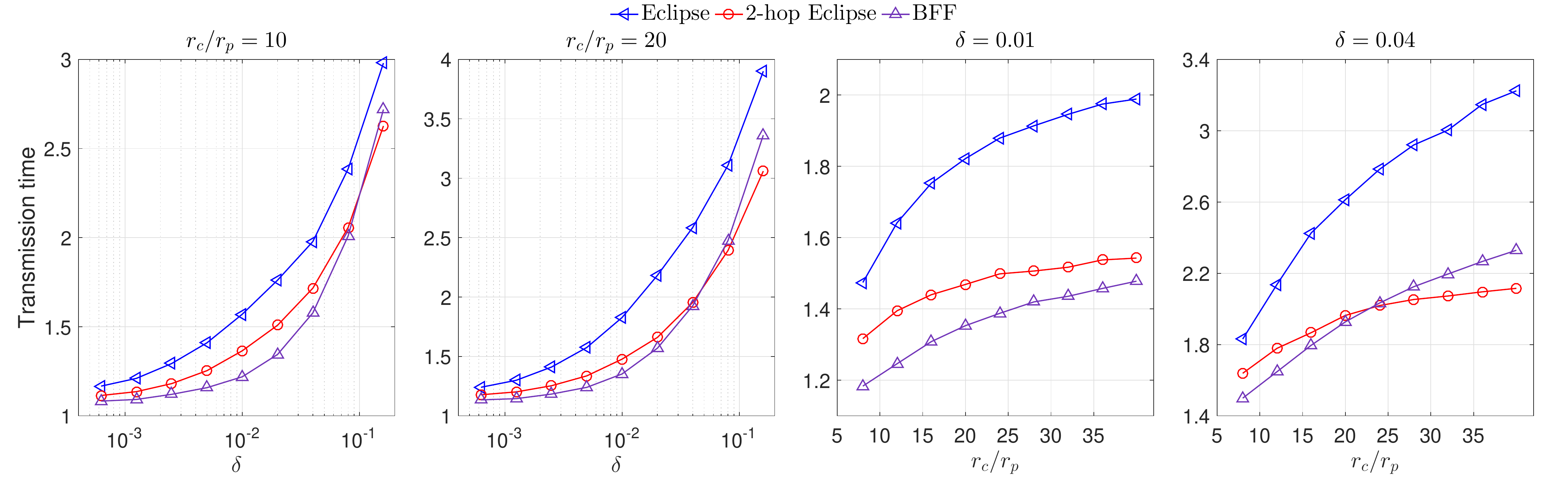}
\caption{Performance comparison of Eclipse, 2-hop Eclipse and BFF under different system setting}\label{fig:figure11}
\end{figure*}

\begin{figure*}
\centering
\includegraphics[width=0.95\textwidth]{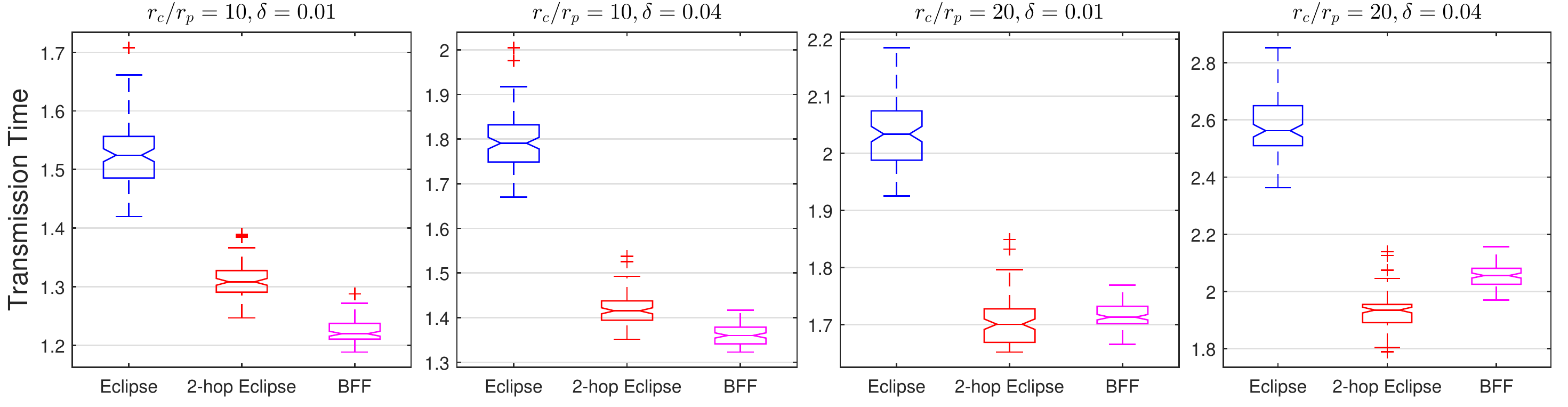}
\caption{Performance comparison of Eclipse, 2-hop Eclipse and BFF under different system setting (boxplot)}\label{fig:boxplot1}
\end{figure*}

\begin{figure*}
\centering
\includegraphics[width=0.95\textwidth]{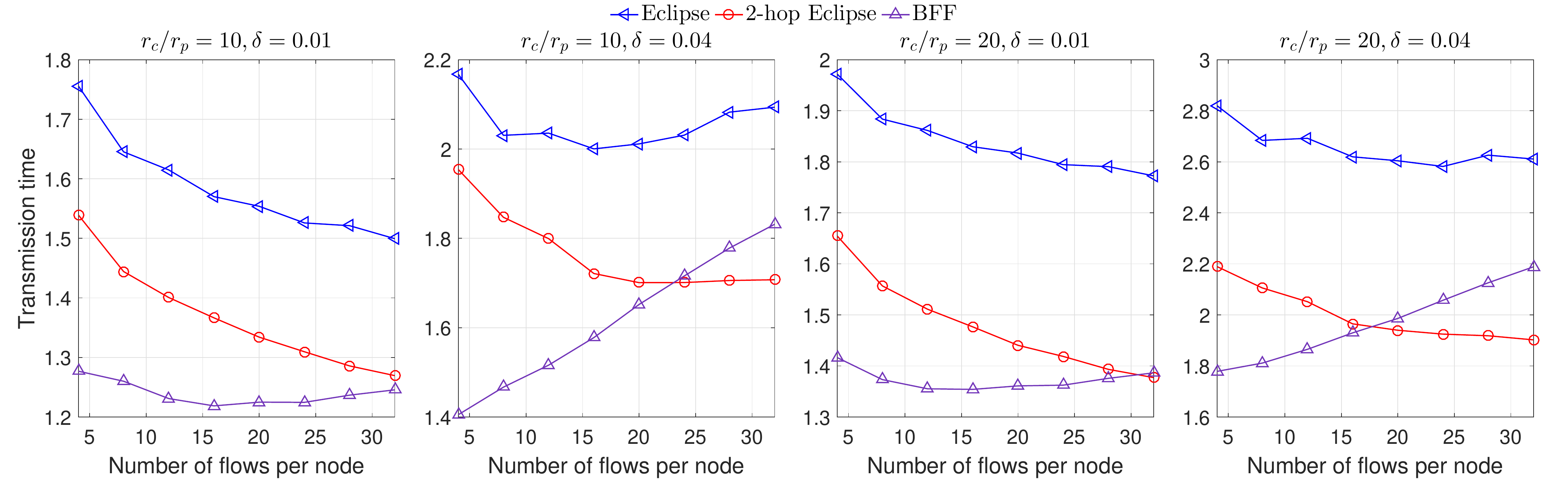}
\caption{Performance comparison of Eclipse, 2-hop Eclipse and BFF while varying sparsity of demand matrix}\label{fig:figure22}
\end{figure*}

\begin{figure*}
\centering
\includegraphics[width=0.95\textwidth]{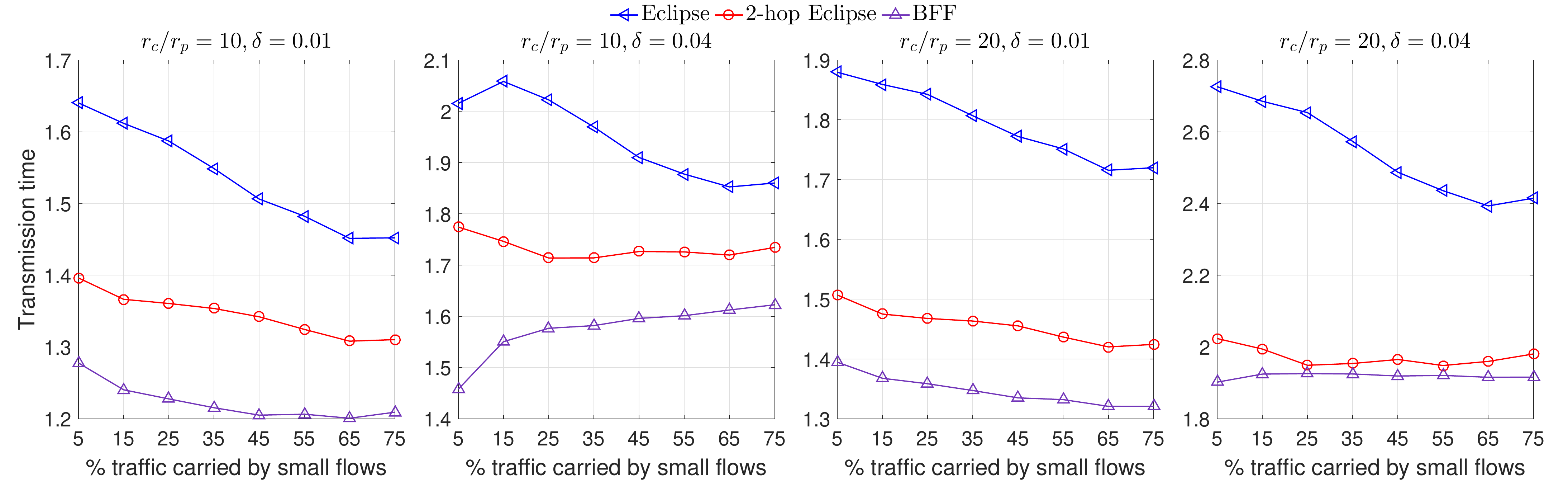}
\caption{Performance comparison of Eclipse, 2-hop Eclipse and BFF while varying skewness of demand matrix}\label{fig:figure33}
\end{figure*}

\subsection{System Parameters}\label{sec:system_setting}

In this section, we introduce the system parameters (of the hybrid switch) used in our simulations.

\textbf{Network size:} We consider the hybrid switch with $n=100$ input/output ports throughout this section.  Other reasonably large (say $\ge 32$) switch sizes
produce similar results.


\textbf{Circuit switch per-port rate $r_c$ and packet switch per-port rate $r_p$:}  As far as designing hybrid switching algorithms is concerned,
only their ratio $r_c/r_p$ matters.
This ratio roughly corresponds to the percentage of traffic that needs to be transmitted by the circuit switch. The higher this ratio is,
the higher percentage of traffic should be transmitted by the circuit switch. This ratio varies from $8$ to $40$ in our simulations.
As explained earlier, we normalize $r_c$ to $1$ throughout this paper.

Since both the traffic demand to each input port and the per-port rate of the circuit switch are all normalized to $1$, the (idealistic) transmission time would be $1$
when there was no packet switch,
the scheduling was perfect (\textit{i.e.}, no ``slack'' anywhere), and there was no reconfiguration penalty (\textit{i.e.}, $\delta = 0$).   Hence we should expect that all these
algorithms result in transmission times larger than $1$ under realistic ``operating conditions'' and parameter settings.

\textbf{Reconfiguration delay (of the circuit switch) $\delta$: } 
In general, the smaller this reconfiguration delay is, the less time the circuit switch has to spend on reconfigurations.
Hence, given a traffic demand matrix, the transmission time should increase as $\delta$ increases. 


\subsection{Performances under Different System Parameters}\label{sec:perform_system_setting}

In this section, we evaluate the performances of Eclipse, 2-hop Eclipse, and BFF for different value combinations of $\delta$ and $r_c/r_p$
under the traffic demand matrix with the default parameter settings ($4$ large flows and $12$ small flows accounting for roughly $70\%$ and $30\%$
of the total traffic demand into each input port).
We perform $100$ simulation runs for each scenario, and report the average values of the simulation results in \autoref{fig:figure11}.
The average results, presented in~\autoref{fig:figure11}, show that the schedules generated by 2-hop Eclipse and BFF are consistently better (\textit{i.e.}, shorter transmission times) than those generated by Eclipse, especially when reconfiguration delay $\delta$ and rate ratio $r_c/r_p$ is large.  More specifically, when $\delta=0.01,r_c/r_p=10$ (default setting), the average transmission time of the schedules generated by 2-hop Eclipse is approximately $13\%$ shorter than that of Eclipse.  That of BFF is about $19\%$ shorter than that of Eclipse.
When $\delta=0.04, r_c/r_p=20$, both 2-hop Eclipse and BFF result in $23\%$ shorter transmission time than Eclipse.


We have also compared, using notched boxplots shown in~\autoref{fig:boxplot1}, the performance variances of these three algorithms as observed from these 100 simulation runs.  
In each notched boxplot, the top and the bottom of each box correspond to the $75^{th}$ and the $25^{th}$ percentiles respectively, 
the line in the box the $50^{th}$ percentile (the median), and the notch the $95\%$ confidence interval of the median.  
We use four different value combinations of the parameters $\delta$ and $r_c/r_p$ in generating these boxplots: (1) $\delta=0.01$, $r_c/r_p=10$; (2) $\delta=0.01$, $r_c/r_p=20$; (3) $\delta=0.04$, $r_c/r_p=10$; and (4) $\delta=0.04$, $r_c/r_p=20$.
As shown in \autoref{fig:boxplot1}, the box of 2-hop Eclipse is thinner than that of Eclipse in every scenario, indicating that a consistently more stable transmission time performance (i.e., lower in variance).  
Also shown in \autoref{fig:boxplot1} is that BFF has the thinnest box in every scenario, indicating its superior performance stability over both Eclipse and 2-hop Eclipse.

\subsection{Performances under Different Traffic Demands}\label{sec:perform_traffic_demand}

In this section, we evaluate the performance robustness of our algorithms 2-hop Eclipse and BFF under a large set of traffic demand matrices
that vary by sparsity and skewness.
We control the sparsity of the traffic demand matrix $D$ by varying the total number of flows ($n_L+n_S$) in each row from $4$ to $32$, while fixing the ratio of the number of large flow to that of
small flows ($n_L/n_S$) at $1:3$.
We control the skewness of $D$ by varying $c_S$, the total percentage of traffic carried by small flows,
from $5\%$ (most skewed as large flows carry the rest $95\%$) to $75\%$ (least skewed).
In all these evaluations, we consider four different value combinations of system parameters
$\delta$ and $r_c/r_p$: (1) $\delta=0.01, r_c/r_p=10$; (2) $\delta=0.01, r_c/r_p=20$; (3) $\delta=0.04, r_c/r_p=10$; and
(4) $\delta=0.04, r_c/r_p=20$.

\autoref{fig:figure22} compares the (average) transmission time of 2-hop Eclipse, BFF, and Eclipse when the sparsity parameter $n_L+n_S$ varies from $4$ to $32$ and the value of the skewness parameter
$c_S$ is fixed at $0.3$.   \autoref{fig:figure33} compares the (average) transmission time of 2-hop Eclipse, BFF, and Eclipse when the the skewness parameter
$c_S$ varies from $5\%$ to $75\%$ and the sparsity parameter $n_L+n_S$ is fixed at $16$ ($=4 + 12$).
In each figure, the four subfigures correspond to the
four value combinations of $\delta$ and $r_c/r_p$ above.



Both \autoref{fig:figure22} and \autoref{fig:figure33} show that 2-hop Eclipse and BFF perform better than Eclipse under various traffic demand matrices.
Furthermore, BFF outperforms 2-hop Eclipse in most cases (the first and third subfigures in \autoref{fig:figure22} and all subfigures in \autoref{fig:figure33}),
demonstrating the benefit that the partial reconfiguration capability brings to BFF.
However, when the traffic demand matrix becomes dense (as the number of flows $n_L+n_S$ increases in \autoref{fig:figure33}), 2-hop Eclipse starts to have an advantage.
As shown in the second and the fourth subfigures of \autoref{fig:figure33} (where the reconfiguration delay is large: $\delta=0.04$),
2-hop Eclipse performs better than BFF.  This shows that 2-hop indirect routing can reduce transmission time significantly under a dense traffic demand matrix.
This is not surprising:  Dense matrix means smaller matrix elements, and it is more likely for a small matrix element to be transmitted entirely by indirect routing
(in which case there is no need to pay a large reconfiguration delay for the direct routing of it) than for a large one.

The boxplots of some simulation results described in this section are shown in \autoref{sec:eval_appendix}.  They all convey the same message as~\autoref{fig:boxplot1} 
that BFF is more stable in transmission time performance than 2-hop Eclipse, which is in turn more stable than Eclipse.

\subsection{Execution time comparison of Eclipse, 2-hop Eclipse and BFF}\label{sec:execution_time}

In this section, we present the execution times of Eclipse, 2-hop Eclipse and BFF algorithms (all implemented in C++) for different $\delta$, under the traffic demand matrix with the default parameter settings ($n_L=4$, $n_S=12$, $c_L=0.7$, $c_S=0.3$). We set $r_c/r_p=10$ for each scenario. 
These execution time measurements are performed on a Dell Precision Tower 3620 workstation equipped with an 
Intel Core i7-6700K CPU @4.00GHz processor and 16GB RAM, and running Windows 10 Professional. We perform $100$ simulation runs for each scenario. The simulation results are shown in \autoref{tab:execution_time}.

\begin{table}[h]
  \setlength{\tabcolsep}{3pt}
  \centering
	\begin{tabular}{@{}c|c|c|c|c|c|c@{}}
	\hline
	\hline
	 & \multicolumn{3}{c|}{$n=32$} & \multicolumn{3}{c}{$n=100$} \\
	\hline
	$\delta$ & $0.0025s$ & $0.01s$ & $0.04s$ & $0.0025s$ & $0.01s$ & $0.04s$ \\
	\hline
	Eclipse & $1.25s$ & $0.80s$ & $0.44s$ & $34.6s$ & $16.4s$ & $6.88s$ \\
	\hline
	2-hop Eclipse & $1.53s$ & $1.02s$ & $0.58s$ & $45.05s$ & $25.91s$ & $11.13s$ \\
	\hline
	BFF & $2.50ms$ & $2.34ms$ & $1.93ms$ & $30.1ms$ & $22.6ms$ & $17.4ms$ \\
	\hline
	\hline
	\end{tabular}
	\caption{Comparison of execution time for Eclipse, 2-hop Eclipse and BFF}
  \label{tab:execution_time}
\end{table}

As shown in~\autoref{tab:execution_time}, the execution time of 2-hop Eclipse is roughly $20\%$ to $40\%$ longer than that of Eclipse, which is consistent with 
fact that each iteration of 2-hop Eclipse is strictly more computationally expensive than each iteration of Eclipse (see \autoref{sec:complexity_twohop}). On the other hand, the execution time of BFF is roughly three orders of magnitude smaller than those of Eclipse and 2-hop Eclipse. We have also implemented Eclipse++ and measured its execution time. It is roughly three orders of magnitude higher than those of Eclipse;  the same observation \cite{personal} was made by the first author of~\cite{bojjacostly} (the Eclipse/Eclipse++ paper).   As explained in~\autoref{sec:parallel_eclipse} and 
shown in~\autoref{tab:execution_time}, the execution times of Eclipse and 2-hop Eclipse are a bit too long but can be shortened via parallelization, which we will elaborate on
in~\autoref{sec:eval_parallel_2hop_eclipse}.

\subsection{Compare 2-hop Eclipse with Eclipse++}\label{sec:evaluation_eclipseplus}

\begin{figure}
\centering
\includegraphics[width=3in]{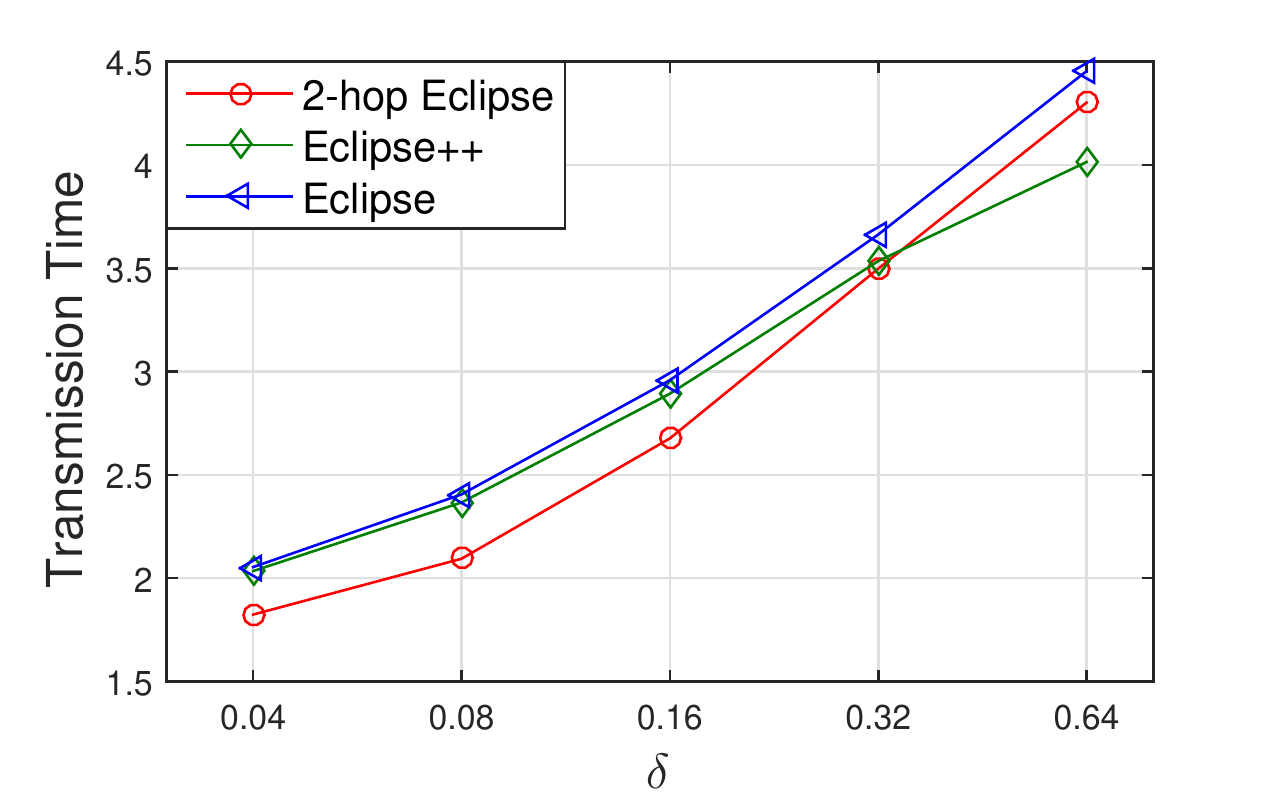}
\caption{Performance comparison of Eclipse, 2-hop Eclipse and Eclipse++}
\label{fig:plusplusvs2hop}
\end{figure}

In this Section, we compare the performances of 2-hop Eclipse and Eclipse++, both indirect routing algorithms, under the default parameter settings.
Since Eclipse++ has a very high computational complexity, we perform only $50$ simulation runs for each scenario.
The results are shown in~\autoref{fig:plusplusvs2hop}.   They show that Eclipse++ slightly outperforms 2-hop Eclipse only when the reconfiguration delay is ridiculously large
($\delta=0.64$ unit of time);  note that, as explained earlier, the idealized transmission time is $1$ (unit of time)!
In all other cases, 2-hop Eclipse performs much better than Eclipse++, and Eclipse++ performs only slightly better than Eclipse.


\subsection{Performance of a Simple Parallel 2-hop Eclipse Scheme}
\label{sec:eval_parallel_2hop_eclipse}

As explained in Section~\ref{sec:parallel_eclipse}, our solution approach to parallelizing 2-hop Eclipse is perform 
MWM computations using only a small subset of parameter values sampled from the space of $n^2$ possible parameter values.
As a proof of concept, we experimented on a simple (and possibly naive) sampling strategy
in which $v_{(m)}$, $v_{(2m)}$, $v_{(3m)}$, $\cdots$, $v_{(\lfloor n^2/m \rfloor\cdot m)}$ 
are sampled where $v_{(1)}$, $v_{(2)}$, $\cdots$, $v_{(n^2)}$ are a permutation of these $n^2$ possible parameter values in the nondecreasing order, and $1/m$ is the sampling rate.  We again use the traffic demand matrix with the default parameter settings ($n_L=4$, $n_S=12$, $c_L=0.7$, $c_S=0.3$). 
Our experimental results, presented in \autoref{tab:interpolation}, show that even when the sampling rate is as small as $\frac{1}{4n}$ (\textit{i.e.}, using only $n/4$ parallel processors), the transmission time performance of the resulting algorithm is only slightly worse (about $8\%$ additional transmission time) than that of the un-sampled algorithm.
When $n = 32$, the parallel 2-hop Eclipse can compute a schedule in tens of milliseconds using only $32/4 = 8$ parallel processors, yet these schedules are
only slightly worse in transmission time performance than the sequential (\textit{i.e.}, not parallelized) 2-hop Eclipse.


\begin{table}[!ht]
\centering
\begin{tabular}{|c|c|c|c|c|c|}
\hline
Sampling rate $(1/m)$ & $1$ & $\frac{1}{n}$ & $\frac{1}{2n}$ & $\frac{1}{4n}$ &$\frac{1}{8n}$  \\
\hline
Transmission Time & $1.215$ & $1.197$ & $1.214$ & $1.318$ & $1.449$ \\
\hline
\end{tabular}
\caption{Transmission time performance with different sampling rates ($n=32$)}
\label{tab:interpolation}
\end{table}

\section{Related Work}
\label{sec: related-work}

In this section, we briefly survey the related work on hybrid switching
other than Eclipse and Eclipse++, which we have thoroughly described and compared
our solutions against earlier.

\subsection{Hybrid Switch Scheduling Algorithms}
\label{subsec: hybrid-switch}

Liu et al.~\cite{liu2015scheduling} first characterized the mathematical problem of the hybrid switch scheduling using direct routing only
and proposed a greedy heuristic solution, called Solstice. In each iteration, Solstice effectively tries to find the Max-Min Weighted Matching (MMWM) in $D$, which is the full matching with the largest minimum element. The duration of this matching (configuration) is then set to this largest minimum element.
The Solstice~\cite{liu2015scheduling} work mentioned the technological feasibility of partial reconfiguration, but made no attempt to exploit this capability.


This hybrid switching problem has also been considered in two other works~\cite{Ghobadi2016projector,Hamedazimi2014FireFly}.  Their problem formulations are a bit different than that in~\cite{bojjacostly,liu2015scheduling}, and so are their solution approaches.  In~\cite{Ghobadi2016projector}, the problem of matching senders with receivers is modeled as a (distributed) stable marriage problem, in which a sender's preference score for a receiver is equal to the age of the data the former has to transmit to 
the latter in a scheduling
epoch, and is solved using a variant of the Gale-Shapely algorithm~\cite{gale1962stablematching}.  This solution is aimed at minimizing transmission latencies while avoiding starvations, and not at maximizing network throughput, or equivalently 
minimizing transmission time. 
The innovations of~\cite{Hamedazimi2014FireFly} are mostly in the aspect of systems building and are not on matching algorithm designs.

To the best of our knowledge, Albedo~\cite{li2017using} is the only other indirect routing solution for hybrid switching, besides Eclipse++~\cite{bojjacostly}.  Albedo was proposed to solved a different
type of hybrid switching problem: dealing with the fallout of inaccurate estimation of the traffic demand matrix $D$.  It works as follows.
Based on an estimation of $D$, Albedo first computes a direct routing schedule using Eclipse or Solstice.  Then any unexpected ``extra workload'' resulting from the inaccurate estimation is routed indirectly.  Just like Eclipse++,
each (indirect routing) iteration of Albedo is to find the shortest indirect routing path (but with a different weight measure than that used in Eclipse++).
However, unlike Eclipse++, where scheduling each input-output flow (\textit{i.e.}, VOQ) requires at least one iteration,
in Albedo, scheduling each TCP/UDP flow belonging to the unexpected ``extra workload'' requires at least one iteration.  Hence we believe the computational
complexity of Albedo, not mentioned in~\cite{li2017using}, is at least as high as that of Eclipse++, if not higher.

\subsection{Optical Switch Scheduling Algorithms}
\label{subsec: optical-switch}



Scheduling of circuit switch alone (\textit{i.e.}, no packet switch), that is not partially reconfigurable, has been studied for decades.
Early works often assumed the reconfiguration delay to be either zero \cite{inukai1979efficient,Porter2013TMS} or infinity \cite{towles2003guaranteed,wu2006nxg05,gopal1985minimizing}. Further studies, like DOUBLE~\cite{towles2003guaranteed}, ADJUST \cite{li2003scheduling} and other algorithms such as \cite{wu2006nxg05,fu2013cost}, take the actual reconfiguration delay into consideration.
Recently, a solution called Adaptive MaxWeight (AMW) \cite{wang2015end,wang2017heavy} was proposed for optical switches (with nonzero reconfiguration delays). The basic idea of AMW is that
when the maximum weighted configuration (matching) has a much higher weight than the current configuration, the optical switch is reconfigured to the maximum weighted configuration;
otherwise, the configuration of the optimal switch stays the same.
However, this algorithm may lead to long queueing delays (for packets) since it usually reconfigures infrequently.

Towles et al. \cite{towles2003guaranteed} first considered the scheduling of circuit switch (alone) that is partially reconfigurable and discovered that such a scheduling problem
is algorithmically equivalent to open-shop scheduling (OSS)~\cite{pinedo2016scheduling}.
They tried to adapt
List scheduling (LIST) \cite{shmoys1994improved,hochba1997approximation,brasel2008heuristic}, the well-known family of polynomial-time heuristic algorithms,
to tackle this problem. However, no algorithm in the LIST family benefits much from the partial reconfiguration capability, as explained earlier.
Recently, Van et al.~\cite{van2017adaptive} proposed a solution called adaptive open-shop algorithm (AOS), for scheduling partially reconfigurable optical switches. It essentially runs an
optimal preemptive strategy~\cite{gonzalez1976open} and a non-preemptive LIST strategy~\cite{pinedo2016scheduling,gonzalez1977bounds} in a dynamic and
flexible fashion to find a good schedule.
However, the computational complexity of the optimal preemptive strategy~\cite{gonzalez1976open} alone is $O(n^4)$, which is much higher than the total complexity of BFF.

\subsection{Related Switching Architectures}

In~\cite{vargaftik2016composite}, Shay et al. extended the classic hybrid switching architecture by introducing one-to-many and many-to-one
composite paths between the circuit switch and the packet switch to improve
the performance of hybrid switch under one-to-many and/or many-to-one traffic workloads.
They proposed a simple technique to transform the problem of scheduling such a special hybrid switch into
that of scheduling a standard hybrid switch.  The transformed problem is then solved using existing solutions such as Solstice~\cite{liu2015scheduling} and Eclipse~\cite{bojjacostly}.
However, no new solutions to the standard hybrid switching problem were proposed in~\cite{vargaftik2016composite}.

\section{Conclusion}\label{sec:conclusion}

We proposed two approaches to significantly reducing transmission time in switch scheduling in the current standard Eclipse algorithm. Our 2-hop Eclipse algorithm achieves the indirect routing improvements found in Eclipse++ while avoiding a significant increase in computation costs.
Our Best First Fit (BFF) algorithm takes advantage of partially reconfigurable circuit switches and achieves similar or better performance but with much lower complexity.


%
%


\bibliographystyle{plain}

\bibliography{load_balancing}
\appendix

\section{Additional boxplots}\label{sec:eval_appendix}

\begin{figure*}[b]
\centering
\includegraphics[width=0.95\textwidth]{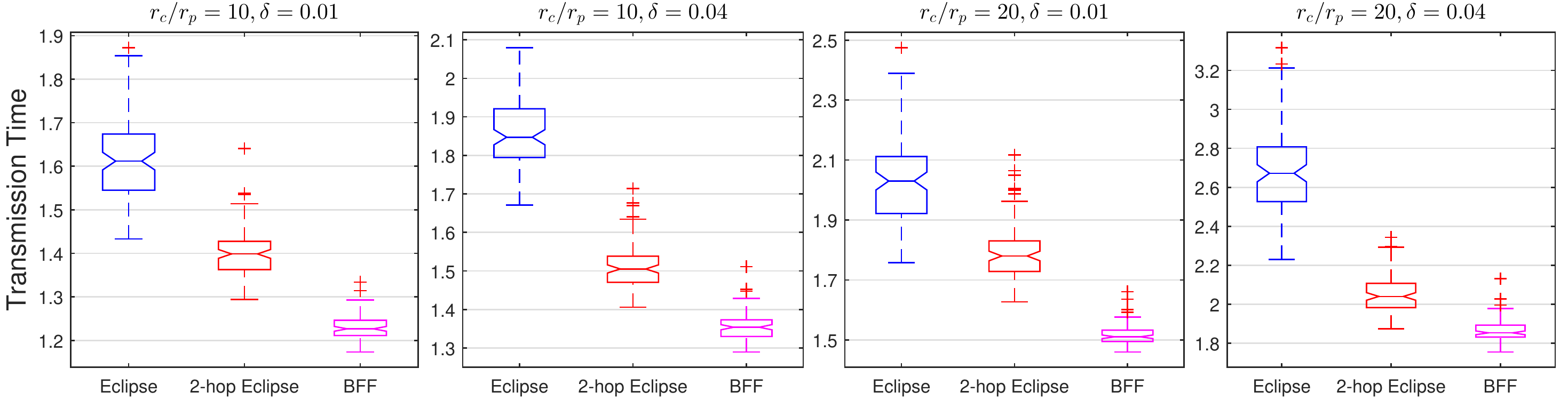}
\caption{Performance comparison of Eclipse, 2-hop Eclipse and BFF under different system setting when $n_l=3, n_s=9$}\label{fig:boxplot2}
\end{figure*}

\begin{figure*}
\centering
\includegraphics[width=0.95\textwidth]{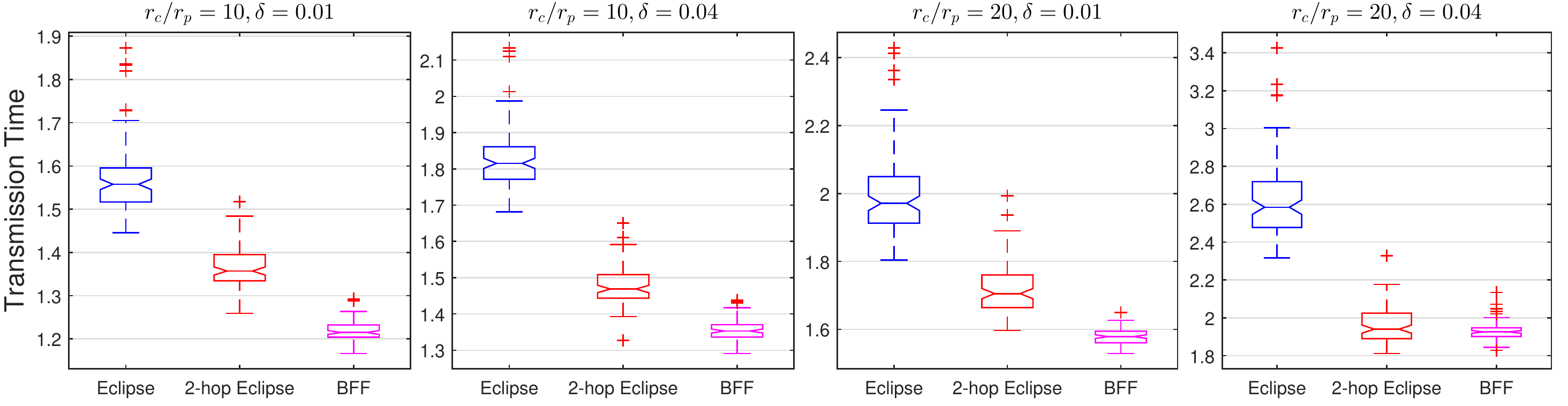}
\caption{Performance comparison of Eclipse, 2-hop Eclipse and BFF under different system setting when $n_l=6, n_s=18$}\label{fig:boxplot3}
\end{figure*}

\begin{figure*}
\centering
\includegraphics[width=0.95\textwidth]{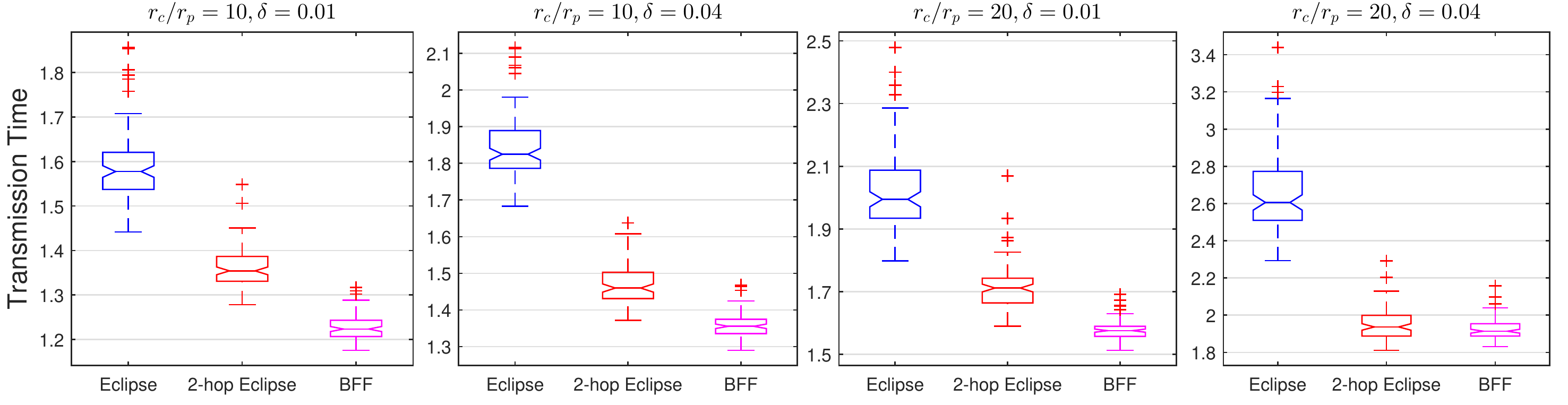}
\caption{Performance comparison of Eclipse, 2-hop Eclipse and BFF under different system setting when $c_l=0.75, c_s=0.25$}\label{fig:boxplot4}
\end{figure*}

\begin{figure*}
\centering
\includegraphics[width=0.95\textwidth]{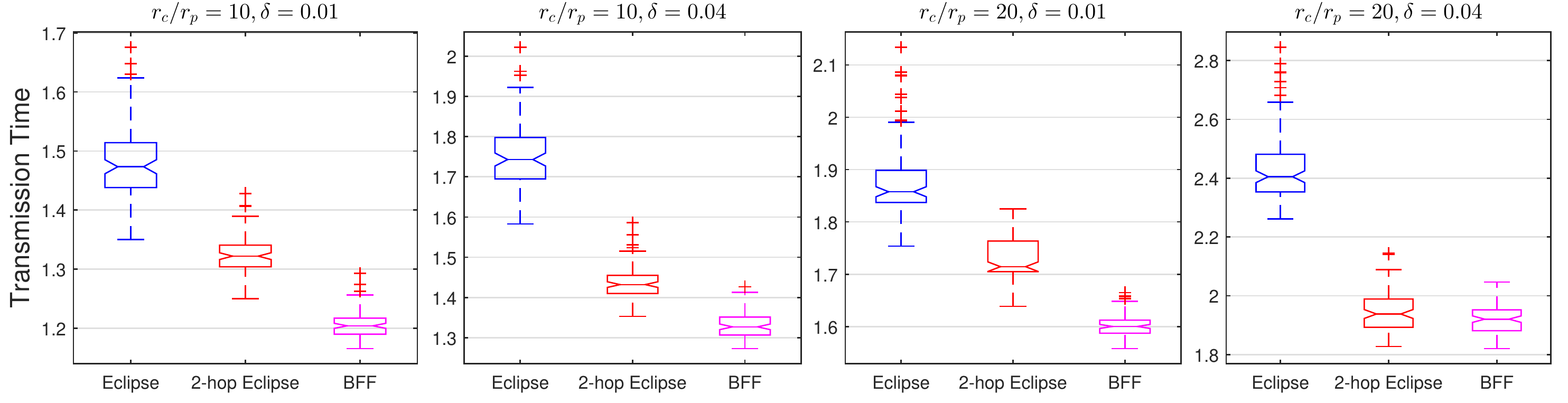}
\caption{Performance comparison of Eclipse, 2-hop Eclipse and BFF under different system setting when $c_l=0.45, c_s=0.55$}\label{fig:boxplot5}
\end{figure*}

In this section, we present the boxplots of some simulation results described in~\autoref{sec:perform_traffic_demand}.  
Recall that in \autoref{sec:perform_traffic_demand}, we control the sparsity of the traffic demand matrix $D$ by varying the total number of flows ($n_L+n_S$), and control the skewness of $D$ by varying $c_S$, the total percentage of traffic carried by small flows.  The simulation results with four different value combinations of sparsity ($n_L+n_S$) and skewness ($c_S$) parameters are plotted in 
Figures~\ref{fig:boxplot2}--\ref{fig:boxplot5} respectively as follows.
In \autoref{fig:boxplot2} and \autoref{fig:boxplot3}, we set the total number of flows ($n_L+n_S$) to $12$ and $24$ respectively, while fixing the ratio of the number of large flow to that of small flows ($n_L:n_S$) 
at $1:3$, and fix the skewness parameter $c_S$ at $0.3$ in these two scenarios;  in \autoref{fig:boxplot4} and \autoref{fig:boxplot5}, we set the skewness parameter $c_S$ to $0.25$ and $0.55$ respectively, while fixing the number of large and small flows as $n_L=4, n_S=12$. In each of these four figures, we consider four different value combinations of system parameters $\delta$ and $r_c/r_p$: (1) $\delta=0.01, r_c/r_p=10$; (2) $\delta=0.01, r_c/r_p=20$; (3) $\delta=0.04, r_c/r_p=10$; and (4) $\delta=0.04, r_c/r_p=20$. 
All four figures convey the same message as~\autoref{fig:boxplot1} 
that BFF is more stable in transmission time performance (has thinner boxes) than 2-hop Eclipse, which is in turn more stable than Eclipse.


\end{document}